\documentclass{article}

\usepackage[utf8]{inputenc}
\usepackage{amsfonts}
\usepackage{natbib}
\usepackage{amssymb}
\usepackage[left=25mm,top=25mm,bottom=25mm,right=25mm]{geometry}
\usepackage{dsfont,amscd}
\usepackage{graphicx}
\usepackage[cyr]{aeguill}
\usepackage{amssymb}
\usepackage{amsmath}
\usepackage{todonotes}
\usepackage{amsthm}
\usepackage{float}
\usepackage{arydshln}
\usepackage{subfig}
\usepackage{xcolor}
\usepackage[all]{xy}
\newcommand{\be}{\begin{equation}}
\newcommand{\ee}{\end{equation}}
\newcommand{\ben}{\begin{equation*}}
\newcommand{\een}{\end{equation*}}
\newcommand{\ban}{\begin{eqnarray*}}
\newcommand{\ean}{\end{eqnarray*}}

\newcommand{\V}[1]{\underline{\mathrm{#1}}}  
\newcommand{\dT}[1]{\underset{\sim}{\mathrm{#1}}}
\newcommand{\tTd}[1]{\underset{\simeq}{\mathrm{#1}}}
\newcommand{\qT}[1]{\underset{\approx}{\mathrm{#1}}} 
\newcommand{\cTd}[1]{\underset{\approxeq}{\mathrm{#1}}}
\newcommand{\sT}[1]{\underset{\underset{\sim}{\approx}}{\mathrm{#1}}}

\newcommand{\ii}{\mathbf{e}_{1}}
\newcommand{\jj}{\mathbf{e}_{2}}

\newcommand{\RR}{\mathbb{R}}

\newcommand{\DD}{\mathrm{D}}

\usepackage{authblk}

\begin{document}

\title{On the validity range of strain-gradient elasticity: a mixed static-dynamic identification procedure}
\author[a]{Giuseppe Rosi}
\author[b]{Luca Placidi}
\author[a]{Nicolas Auffray}

\affil[a]{Université Paris-Est, Laboratoire Modélisation et Simulation Multi Échelle, 
MSME UMR 8208 CNRS, 61, avenue du général de Gaulle, 94010 Créteil Cedex, France} 
\affil[b]{Faculty of Engineering, International Telematic University Uninettuno, Rome, Italy}
\affil[c]{Université Paris-Est, Laboratoire Modélisation et Simulation Multi Échelle, 
MSME UMR 8208 CNRS, 5, Boulevard Descartes, 77454 Marne La Vallée, France} 
\maketitle

\begin{abstract}
Wave propagation in architectured materials, or materials with microstructure, is known to be dependent on the ratio between the wavelength and a characteristic size of the microstructure. Indeed, when this ratio decreases (i.e. when the wavelength approaches this characteristic size) important quantities, such as phase and group velocity, deviate considerably from their low frequency/long wavelength values. This well-known phenomenon is called dispersion of waves.
Objective of this work is to show that strain-gradient elasticity can be used to quantitatively describe the behaviour of a microstructured solid, and that the validity domain (in terms of frequency and wavelength) of this model is sufficiently large to be useful in practical applications.
To this end, the parameters of the overall continuum are identified for a periodic architectured material, and the results of a transient problem are compared to those obtained from a finite element full field computation on the real geometry. 
The quality of the overall description using a strain-gradient elastic continuum is compared to the classical homogenization procedure that uses Cauchy continuum. The extended model of elasticity is shown to provide a good approximation of the real solution over a wider frequency range.
\end{abstract}

\section{Introduction}\label{sec:intro}

\begin{figure}[H]
\centering
\includegraphics[scale=0.5]{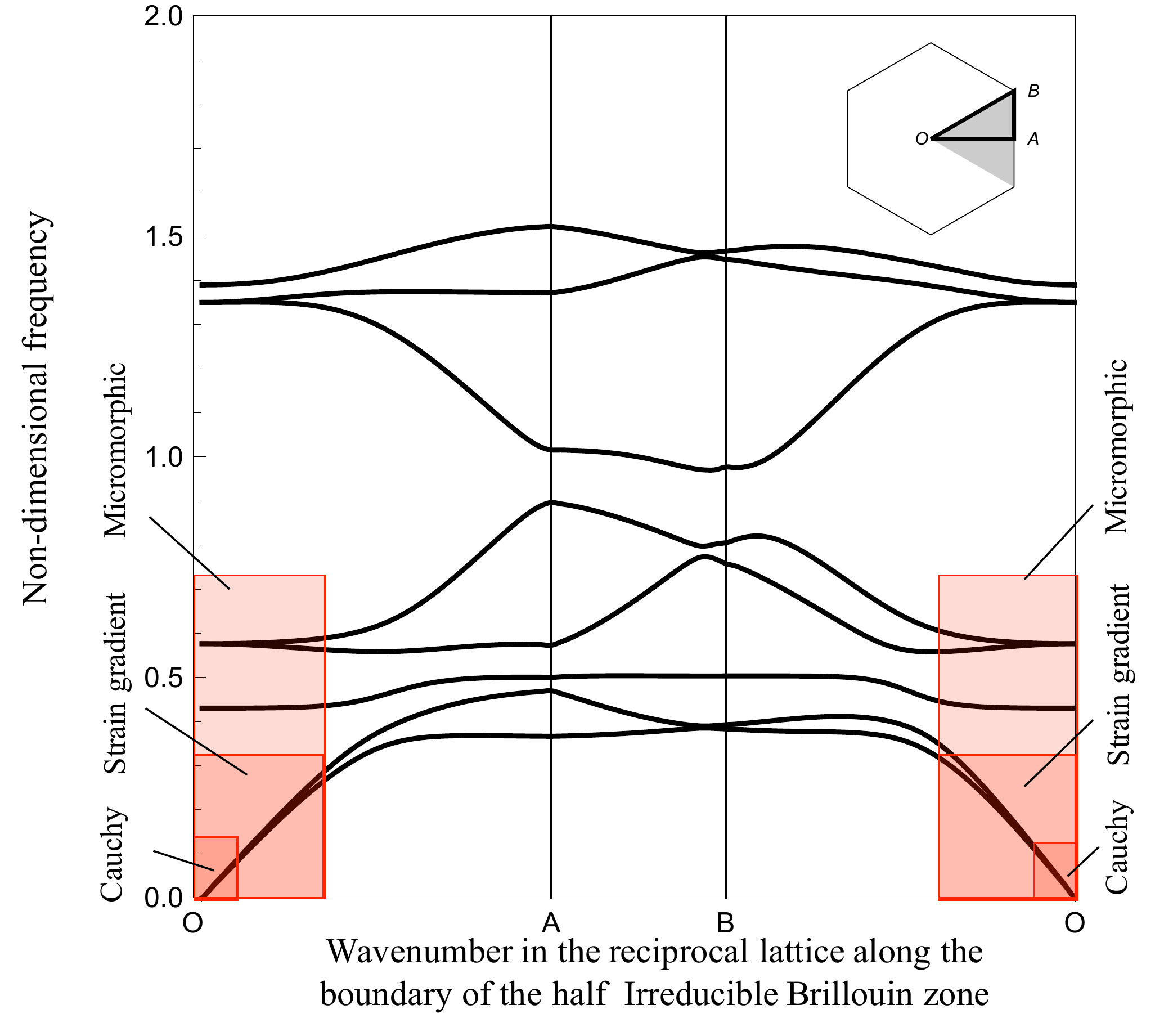}\protect\caption{\label{Fig3}Characteristic dispersion curve of a periodic architectured material (Image extracted from \cite{RA16}). }
\label{Fig.DispExa}
\end{figure}

The description of the wave propagation in a medium having an inner architecture poses a methodological problem. One is facing the following alternative:
either the internal architecture is "infinitely" small with respect to the wavelengths of the solicitation, or it is not.
In the first case, the effects linked to the internal structure are negligible and the architectured material can be replaced by an equivalent standard Cauchy elastic medium\footnote{By Cauchy continuum we simply mean the classical formulation of elasticity.}.
In the second case, structural effects can not be neglected and all the geometrical details of the architecture must be taken into account for computing the wave propagation. The numerical cost of this last option can be prohibitively high. Structural effects related to heterogeneous wave propagation are well illustrated on dispersion diagram. Example of such diagram for an hexagonal lattice material is provided on Fig.\ref{Fig.DispExa}.
Characteristics of wave propagation in heterogeneous continuum are:
\begin{itemize}
\item optical branches;
\item dispersivity;
\item directivity.
\end{itemize}
For practical applications, an intermediate strategy would be valuable. Such a way would consist in defining an equivalent elastic medium retaining certain features of the heterogeneous propagation. But, as well-known, these features can not be captured by the classical theory of Cauchy elasticity.
This observation, among others, motivated, during the 1960s, the construction and study of enriched continuous media, also known as generalized continua \citep{Min64,ME68}. 
As commented in \cite{dell2015origins}, the research on generalized continua is not new and dates back to 1848 with Gabrio Piola, at the origin of continuum mechanics \cite{dell2014complete}.
There are many options for extending classical elasticity and  picking a model among another depends on the nature of the architectural effects that are desired to be maintained at the continuous scale. This choice is application dependent.

In full generality, overall generalized media can either be local or non-local. Despite their interests related to Willis elastodynamics  and cloaking theory, non-local aspects will not be discussed here \citep{Wil85,Wil97,NS11,NHA15a,NHA15b,NHA15c}. Concerning local continua, there are two approaches to extend classical elasticity \citep{Tou62,Min64,Min65,Eri68,ME68}:
\begin{description}
\item[Higher-order continua:] the number of degrees of freedom is extended. The Cosserat model (also known as micropolar), in which local rotations are added as degrees of freedom, belongs to this family \citep{Cos09}. This enhancement can be extended further to obtain the micromorphic elasticity \citep{GR64,Min64,Eri68,Ger73}. This approach allows optical branches to be described.
\item[Higher-grade continua:] the degrees of freedom are kept identical but higher-order gradients of the displacement field are involved into the elastic energy. Within this framework dispersivity and directivity can be described but not optical branches. Strain-gradient elasticity \citep{Min64,Min65,ME68} belongs to this family. It is worth to note that strain-gradient elasticity can be retrieved as a Low Frequency (LF), Long Wave-length (LW) approximation of the micromorphic kinematic \citep{Min64}. As a consequence, the parameters needed to set up this model are limited compared to a complete micromorphic continuum.
\end{description}

The domains of validity of these extended theories are roughly estimated in Fig.\ref{Fig.DispExa}, where it can be observed that in higher-grade continua, due to the absence of internal degrees of freedom, all optical branches are lost. 
Besides, in LW limit, the dispersion relation becomes linear, and hence dispersive effects vanishes.
It should be emphasized that the use of a local generalized continuum provides a good description of the local dynamics only for a short window of wavelengths. 
Formulated differently, there will always be a limit beyond which the substitution medium, as rich as it is, will fail to accurately describe the real dynamics\footnote{Formulated in a third way, local continua can not describe accurately the entire first Brillouin zone, to achieve such a goal non-local continua should be used.}. 
But, and despite of its importance with respect to practical applications, the precise value of this limit is rather unclear.

In the present paper, and following some previous works \citep{ADR15,RA16,PAD+15,PAG16}, attention will be focused on Strain-Gradient Elasticity\footnote{For other modeling options devoted to the description of band-gaps, the reader can refer, among others, to \cite{LHH12,CLH14,NGM14}.} (SGE). Our goal is to define criteria to assess the validity range of the model. The associate procedure is then applied to a material having a square mesostructure. 
If, in the context of this paper, the approach is numerical, it is worth mentioning that the procedure can be applied experimentally.\\

\noindent\textbf{Organization of the paper:}\\ The paper is organized as follows. 
In a first time, \S.\ref{S:EqStrGrd}, the basic equations of strain-gradient elasticity are recapped. 
In \S.\ref{s:QuaAss} a general identification procedure is introduced. 
Then, in  \S.\ref{s:casestudy} this procedure is conducted in the particular case of a tetragonal lattice. The strain-gradient elasticity model is evaluated for this specific situation. Finally, \S.\ref{s:Con} is devoted to some conclusions.\\

\noindent\textbf{Notations:}\\ In this work  tensors of order ranking from $0$ to $6$ are denoted, respectively, by $\mathrm{a}$, $\V{a}$, $\dT{a}$, $\tTd{a}$, $\qT{a}$, $\cTd{a}$ and $\sT{a}$. The simple, double  and fourth contractions are written $\, ., \, :$ and $::$ respectively. 
In index form, with respect to an orthonormal Cartesian basis, these notations correspond to:
\ben
\V{a} . \V{b} = a_i b_i, \quad \dT{a} \, : \, \dT{b} = a_{ij} b_{ij}, \quad \qT{a} \, :: \, \qT{b} = a_{ijkl} b_{ijkl},\qquad 1\leq i,j,k,l\leq d
\een
where repeated indices are summed up. Spatial gradient will classically be denoted, in index form, by a comma: 
\ben
\mathrm{Grad}\ \V{a}=\left(\V{a}\otimes\V{\nabla}\right)_{ij}=a_{i,j}
\een
When needed index symmetries are expressed as follows:  $(..)$  indicates invariance under permutations of the indices in parentheses, while $\underline{..}\ \underline{..}$ denotes invariance with respect to permutations of the underlined blocks.   Finally, a superimposed dot will denote a partial time derivative.

\section{Strain-gradient elasticity in a nutshell}\label{S:EqStrGrd}

In this section  equations of strain-gradient elasticity  are recalled. To that aim the setting introduced by Mindlin (type II formulation) \citep{ME68} is used.

\subsection{Energy}

As usual in field theory of conservative system, the Lagrangian density $\mathcal{L}$ is defined as the difference between the kinetic and potential energy densities, respectively, $\mathcal{K}$ and $\mathcal{P}$.
\ben
\mathcal{L}=\mathcal{K}-\mathcal{P}
\een
In the case of Mindlin's strain gradient theory those quantities are function of the displacement and its gradients as follows:
\begin{align}
\mathcal{K}=\frac{1}{2}p_i v_i+\frac{1}{2}q_{ij} v_{i,j},\qquad
\mathcal{P}=\frac{1}{2}\sigma_{ij}\varepsilon_{ij}+\frac{1}{2}\tau_{ijk}\eta_{ijk}.
\end{align}
The following quantities are involved in these definitions:
\begin{itemize}
\item $p_{i}$ and $q_{ij}$,  the momentum and the hypermomentum tensors;
\item  $v_{i}$ and $v_{i,j}$,  the velocity ($v_{i}=\dot{u}_{i}$) and its gradient;
\item  $\sigma_{ij}$ and $\tau_{ijk}$, the Cauchy stress and the hyperstress tensors;
\item  $\varepsilon_{ij}$ and $\eta_{ijk}=\varepsilon_{ij,k}$, the infinitesimal strain tensor ($\varepsilon_{ij}=(u_{i,j}+u_{j,i})/2$) and its gradient.
\end{itemize}
From the static quantities we can  define the following \emph{total} quantities:
\begin{itemize}
\item 
the total stress
\be
s_{ij}=\sigma_{ij}-\tau_{ijk,k} \label{eq:effectivestress}
\ee
\item 
the total momentum
\be
\pi_{i}=p_{i}-q_{ik,k} \label{eq:effectivemomentum}
\ee
\end{itemize}
 
This form is postulated here on phenomenological basis following \cite{Min64}. It can be noted that the enrichment in the definition of the kinetic energy he introduced in this work has been discarded in its following papers \citep{ME68}. Higher inertia terms were indeed proved to be necessary in more recent publications \citep{AA06,BEB11}, and can be justified by direct asymptotic homogenization approaches \citep{BG14b}, or by localizing Willis equation \citep{NHA15b}.
It can be observed that this Lagrangian is of order one in time and two in space, hence introducing space-time asymmetry \citep{Met06}. Despite of its interest,  the consequence of this observation will not be discussed hereafter.

By application of the least action principle on the action functional \citep{Min64,ME68}, and using the total static quantities previously defined, the following bulk equations are obtained
\begin{equation}
s_{ij,j}+f_{i}=\dot{\pi}_i\label{eq:Bulk}
\end{equation}
Bulk equations are supplemented with the boundary conditions on edges:
\begin{equation}\label{BC_1_2}
\begin{cases}
t_i=(s_{ij}+\dot{q}_{ij})n_{j}-P_{ml}(P_{mj}\tau_{ijk}n_{k})_{,l}\\
R_i=\tau_{ijk}n_{j}n_{k}
\end{cases}
\end{equation}
and on vertexes
\begin{equation}\label{BC_33}
\nu_i=[[\tau_{ijk}n_{j}m_{k}]]
\end{equation}
where the quantities $\V{t},\V{R},\V{\nu}$, $\V{n}$ and $\V{m}$ are, respectively, the traction (i.e. a force per unit length), the double-force per unit length, the vertex-force, the outward normal and the outward tangent.
It is a matter of fact that on each vertex, we have two edges and therefore two outward normals and two outward tangents; the symbol $[[\cdot]]$ means that the quantity $\cdot$ is evaluated first on one edge, then on the other edge and then the sum of the two quantities is calculated.
The quantity $\dT{P}$, which is the projector onto the tangent plane, is defined as follows:
\ben
\dT{P}=\dT{I}-\V{n}\otimes\V{n}
\een
The boundary conditions (\ref{BC_1_2})$_1$ and (\ref{BC_33}) are the dual of the displacement $\V{u}$ and the boundary condition (\ref{BC_1_2})$_2$ is the dual of the normal displacement gradient $\nabla \V{u} \cdot \V{n}$. Thus, a well-posed boundary value problem is given once displacement and normal displacement gradient (or their duals) are imposed at the boundary.

\subsection{Constitutive equations} 
For the mechanical model to be closed, constitutive equations relating primal and dual quantities are mandatory.  
In the present situation, those relations will assumed to have the following structure:
\begin{equation}
\begin{pmatrix}
\V{p}\\
\dT{q}\\
\dT{\sigma}\\
\tTd{\tau}
\end{pmatrix}
=
\begin{pmatrix}
\rho\dT{I}&\tTd{K}&0&0 \\
\tTd{K}^{T}&\qT{J}&0&0\\
0&0&\qT{C}&\cTd{M}\\
0&0&\cTd{M}^T&\sT{A}
\end{pmatrix}
\begin{pmatrix}
\V{v}\\
\dT{\nabla v}\\
\dT{\varepsilon}\\
\tTd{\eta}
\end{pmatrix}\label{eq:ConstMatrix}
\end{equation}
where
\begin{itemize}
\item $\rho I_{(ij)}$ is the macroscopic mass density;
\item $K_{ijk}$ is the coupling inertia tensor;
\item $J_{ijqr}$ is the second order inertia tensor.
\item $C_{\underline{(ij)}\  \underline{(lm)}}$ is the classical elasticity tensor;
\item $M_{(ij)(lm)n}$ is a fifth-order coupling elasticity tensor;
\item $A_{\underline{(ij)k}\ \underline{(lm)n}}$ a six-order tensor. 
\end{itemize}
In the case of centrosymmetric continuum\footnote{A periodic lattice is said to be centrosymmetric if its unit cell is invariant under the inversion operation $(-\dT{I} \in \mathrm{O}(2))$. In 2D this is equivalent for the unit cell to be invariant with respect to a rotation of angle $\pi$ \citep{OA14}}, odd-order tensors vanish and Eq.\eqref{eq:ConstMatrix} simplifies to
\begin{equation}
\begin{pmatrix}
\V{p}\\
\dT{q}\\
\dT{\sigma}\\
\tTd{\tau}
\end{pmatrix}
=
\begin{pmatrix}
\rho\dT{I}&0&0&0 \\
0&\qT{J}&0&0\\
0&0&\qT{C}&0\\
0&0&0&\sT{A}
\end{pmatrix}
\begin{pmatrix}
\V{v}\\
\dT{\nabla v}\\
\dT{\varepsilon}\\
\tTd{\eta}
\end{pmatrix}\label{eq:ConstMatrix_CS}
\end{equation}
Centrosymmetry will be assumed for the rest of the paper. In 2D space, this assumption is not too restrictive since $\cTd{M}$ and $\tTd{K}$ are null in many common situations \citep{ADR15,AKO16}.
%
%
The substitution of the constitutive equations (\ref{eq:ConstMatrix_CS}) into Eq.\eqref{eq:effectivestress} and Eq.\eqref{eq:effectivemomentum} gives:
\ban
s_{ij}&=&C_{ijlm}\varepsilon_{lm}-A_{ijklmn}\varepsilon_{lm,kn},\\
\pi_{i}&=&\rho v_{i}-J_{ipqr}v_{q,pr}.
\ean
Hence, for null body force, the bulk equilibrium \eqref{eq:Bulk} expressed in terms of the displacement field yields
\begin{equation}
C_{ijlm}u_{l,jm}-A_{ijklmn}u_{l,jkmn}=
\rho \ddot{u}_i-J_{ipqr}\ddot{u}_{q,pr}.
\label{eq:Bulk2}
\end{equation}
This expression will now be used to introduce a generalized acoustic tensor.

\subsection{Plane wave solution and generalized acoustic tensor}

To obtain the different velocities of a plane wave in the framework of strain-gradient elasticity, let us consider the following plane wave solution:
\begin{equation}
u_{i}=U_{i} \mathcal{A} \exp\left[\imath\left(\omega t-k_{i}x_{i}\right)\right]\label{eq:WaveSol1}
\end{equation}
where $\imath$ denotes the imaginary unit, $\omega$ the angular frequency and $\V{k}$  the wave vector. Moreover, $U_i$ is a real valued unitary vector representing the polarization (direction of motion) and $\mathcal{A}$ is a complex amplitude. These quantities are both independent of $x_i$ and $t$.
The wave vector can be also expressed:
\begin{equation}
k_{i}=\frac{\omega}{V}\hat{\xi}_{i}.
\end{equation}
where $V=\left\|\V{v}^{p}\right\|=\left\|\dot{u}\right\|$ is the norm of the phase velocity of the wave-front, $\V{\hat{\xi}}$ the unit vector pointing toward the direction of propagation, i.e. the normal to the wave-front. 
The relation \eqref{eq:WaveSol1} can be rewritten in the following form
\begin{equation}
u_{i}=U_{i} \mathcal{A} \exp\left[\imath\omega\left(t-\dfrac{1}{V}\hat{\xi}_{i}x_{i}\right)\right]\label{eq:WaveSol2}
\end{equation}
The substitution of this ansatz \eqref{eq:WaveSol2} into the balance equation \eqref{eq:Bulk2} yields
\begin{equation}
\left(\left (C_{ijlm} -\omega^{2} J_{ijlm}\right )\hat{\xi}_{j}\hat{\xi}_{m}
+\frac{\omega^{2}}{V^{2}}A_{ijklmn}\hat{\xi}_{j}\hat{\xi}_{k}\hat{\xi}_{m}\hat{\xi}_{n}\right)U_{l}=\rho V^{2} U_{i},
\end{equation}
which can be conveniently rewritten as
\begin{equation}
\hat{Q}_{il}U_{l}=\rho V^{2}U_{i},\label{eq:EVP}
\end{equation}
where the generalized acoustic tensor $\hat{Q}_{il}$ is defined as follows:
\begin{equation}
\hat{Q}_{il}=\left (C_{ijlm} -\omega^{2} J_{ijlm}\right )\hat{\xi}_{j}\hat{\xi}_{m}
+\frac{\omega^{2}}{V^{2}}A_{ijklmn}\hat{\xi}_{j}\hat{\xi}_{k}\hat{\xi}_{m}\hat{\xi}_{n}. \label{eq:BulkPlane}
\end{equation}
As can be noticed, the classic definition of the acoustic tensor is retrieved (i.~e. $Q_{il}=C_{ijlm}\hat{\xi}_{j}\hat{\xi}_{m}$) in the following situations:
\begin{itemize}
\item when the tensors $A_{ijklmn}$ and  $J_{ijlm}$ vanish, that is for a classic continuum;
\item when $\omega\rightarrow 0$, that is for low frequencies.
\end{itemize}
It can further be observed that, since the expression of the generalized acoustic tensor $\hat{Q}_{il}$ is quadratic in $\omega$, it admits a \emph{horizontal} tangent at the origin ($\omega=0$). This remark  has two implications:
\begin{enumerate}
\item It allows the Cauchy elasticity model to be valid in a neighbourhood of $\omega=0$. In case of a linear dependence, this domain would have been restricted to a single point;
\item It gives information on the initial tangent of derived quantities like the phase and the group velocity. Such information is important for curve fitting perspective.
\end{enumerate}
From the solution of the eigenvalue problem associated to Eqn.\eqref{eq:EVP}, it is possible to obtain useful information concerning phase velocity and polarization of plane waves propagating with a wavefront perpendicular to a given direction $\V{\hat{\xi}}$.
Another important quantity is the group velocity, which is defined as 

\begin{equation}
	\V{v}^{g}=\dfrac{\partial \omega}{\partial \V{k}}.
\end{equation}
From equation \eqref{eq:BulkPlane} it can be shown that
\begin{equation}
v^{g}_j=\frac{Q_{ijl}^{\sharp} U_{l} U_{i}}
{V\rho_{ik}^{\sharp}U_{k}U_{i} }\label{eq:GrVel}
\end{equation}
where
\ben 
Q_{ijl}^{\sharp}=\left (C_{ijlm}-\omega^{2}J_{ijlm} \right )\hat{\xi}_{m}+\frac{\omega^{2}}{V^{2}}A^{\sharp}_{ijklmn}\hat{\xi}_{k}\hat{\xi}_{m}\hat{\xi}_{n},\quad\text{with}\ 
A^{\sharp}_{ijklmn}=\left(A_{ikjlmn}+A_{ijklmn}\right)
\een
and
\ben 
\rho_{ik}^{\sharp}=\rho\delta_{ik}+\frac{\omega^{2}}{V^{2}}J_{ijkl}\hat{\xi}_{j}\hat{\xi}_{l}.
\een
As it can be verified from \eqref{eq:GrVel}, group velocity depends explicitly on the polarization vector.

\section{General identification procedure}\label{s:QuaAss}

Now that the strain gradient elasticity has been presented, we aim at introducing a procedure to estimate its quality as a substitution continuum.  To provide a good overall description of true wave propagation, quantities of interest such as phase and group velocities should be correctly described\footnote{The comparison is not only be made on the dispersion curve, but also on the phase and group velocity. The reason is that the domain of validity could be smaller for group velocity rather than for the dispersion curve. This can be explained by the fact that a good description of the dispersion curve does not imply that the associated mode are well described. This point has been demonstrated in the context of Willis equation by \cite{NHA15a}.}. Hence the quality of strain-gradient model will be evaluated by comparing these quantities with their exact values obtained by a Bloch analysis conducted on a periodic cell. 
Our procedure involves the following steps:

\begin{enumerate}
\item Computation of the Bloch solution over a periodic cell of the real structure:
\begin{enumerate}
\item Plot of the dispersion diagram; 
\item Determination of the phase and group velocity curves.
\end{enumerate}
\item Evaluation of the SGE parameters:
\begin{enumerate}
\item Static identification of the elastic tensors $\qT{C}$ and  $\sT{A}$ using numerical experiments;
\item Dynamic identification of the micro inertia tensor $\qT{J}$ using results of Bloch analysis.
\end{enumerate}
\item Evaluation of the discrepancy between the SGE model and the complete one with respect to the wave number $\V{k}$.
\end{enumerate}

Let us detail the dynamic part of the identification processes, which is based on the computation of the dispersion curves for the unitary cell by using Bloch analysis \citep{DDJ08,FL11,GDK+13}.

For the sake of simplicity, and without losing generality, attention will be restricted for the rest of the paper on unidirectional wave propagation.  This case corresponds to a plane wave propagating towards a specific direction, that we suppose fixed. 

Let us denote by $\V{\hat{\xi}}$ this fixed direction,  the wave vector is a vector field  along this direction:
\ben
\V{k}_{(i)}=k_{(i)}\V{\hat{\xi}}
\een
with $k_{(i)}$, the wave number.  For our need, this function will be sampled in the first Brillouin zone, hence provided a discrete set of wavenumbers:
\ben
k_{(i)}=\dfrac{i}{N_p-1} \dfrac{\pi}{a}\qquad\text{for}\qquad i=0,...,N_p-1
\een
where $a$ is the size of the unit cell and $N_p$ the number of points used in the discretization of the first Brillouin zone \citep{Bri03}.

The corresponding angular frequencies are denoted $\omega_n(k_{(i)})$. Since we are only interested in the acoustic branches, and we are in 2D, $n=1,2$.
From this, values of the phase velocities for the first and the second mode can be computed. Referring to the low frequency identification, the first mode is denoted as S- while the the second as P-\footnote{For anisotropic continuum, in a generic direction, modes are neither pure S- nor pure P-. Hence, this notation is a bit abusive since, but consistent with the case study in section \ref{s:casestudy}}, so that:
\ben
\widehat{V}_S(k_{(i)})=\dfrac{\omega_1(k_{(i)})}{k_{(i)}},\qquad \widehat{V}_P(k_{(i)})=\dfrac{\omega_2(k_{(i)})}{k_{(i)}}
\een
where we used the notation $\widehat{\cdot}$ for quantities computed from Bloch analysis.
Next, by using a finite difference approximation of the first derivative, we can compute the group velocities:
\ben
\widehat{v}^{g}_{S}(k_{(i)})=\frac{\omega_1(k_{(i+1)})-\omega_1(k_{(i)})}{k_{(i+1)}-k_{(i)}},\qquad 
\widehat{v}^{g}_{P}(k_{(i)})=\frac{\omega_2(k_{(i+1)})-\omega_2(k_{(i)})}{k_{(i+1)}-k_{(i)}}.
\een
These values will be compared with those obtained from the solution of the eigenvalue problem \eqref{eq:EVP} and from \eqref{eq:GrVel}.

In the identification procedure, it is crucial to choose the correct quantity for performing the fitting. Indeed, three choices are possible: i) dispersion curves; ii) phase velocity; iii) group velocity. Since group velocity is obtained from the derivative of the dispersion curve, it is reasonable to consider that this will be the first quantity to deviate when increasing the wavenumber. Then, group velocity will be used in the fitting procedure, that involves the following minimization:

\be
J^{Opt}_P=\arg \min_{J_P\in\mathbb{R}^+} \sum_{i=1}^{N_l}\left(v_{P}^{g}(k_{(i)},J_P)-\widehat{v}^{g}_{P}(k_{(i)}) \right)^2,\quad
J^{Opt}_S=\arg \min_{J_P\in\mathbb{R}^+} \sum _{i=1}^{N_l}\left(v_{S}^{g}(k_{(i)},J_S)-\widehat{v}^{g}_{S}(k_{(i)}) \right)^2
\label{eq:opt}
\ee
where  $N_l$ corresponds to a given discrete limit value for the wavenumber and $v_P^{g}(k,J_P)$ and $v_S^{g}(k,J_P)$ are group velocities solution of the eigenvalue problem \eqref{eq:EVP}.
The choice of $N_l$ deserves particular attention, and its introduction is based on the following remarks:\
\begin{itemize}
\item a dispersion curve computed for a generalized continuum cannot fit the dispersion curve on the whole first Brillouin zone
\item more weight must be given to points corresponding to low wavenumbers, to ensure a good continuity with the static model.  
\end{itemize}
In this paper, we fixed this limit to that $N_l$ corresponding to the one that maximises the validity range.
Such a validity range is defined as the region of the $k$-axis for which the error in the fitting is less than 1\%.

This procedure will now be applied on a specific situation, that is the objective of the section \ref{s:casestudy}.

\section{A case study: square microstructure}\label{s:casestudy}

\begin{figure}[H]
\centering
\includegraphics[scale=1]{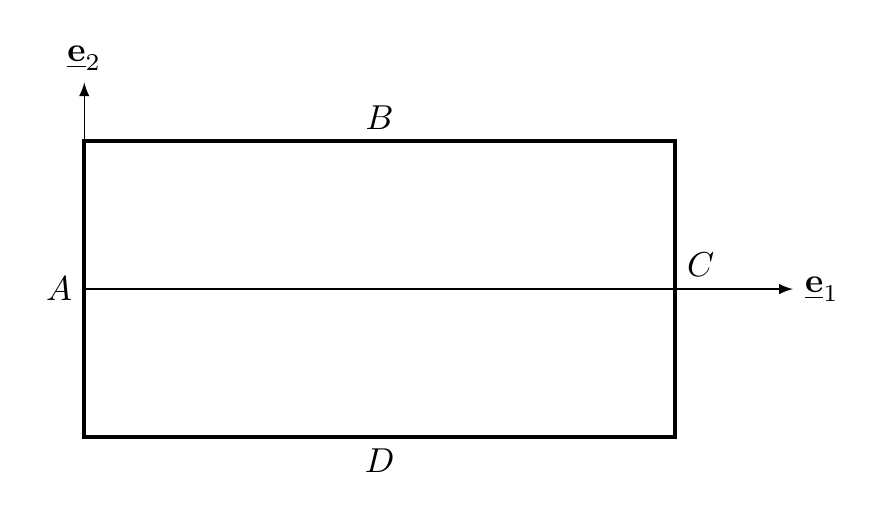}\protect\caption{\label{bodyB}Nomenclature of the 2-dimensional domain.}
\end{figure}

Let us begin by fixing an orthonormal base $\mathcal{B}=(\ii;\jj)$ of $\RR^2$. In this section, a rectangular shape domain will be considered, as depicted in figure Fig.\ref{bodyB}.

\begin{figure}[H]
\centering
\includegraphics[scale=1]{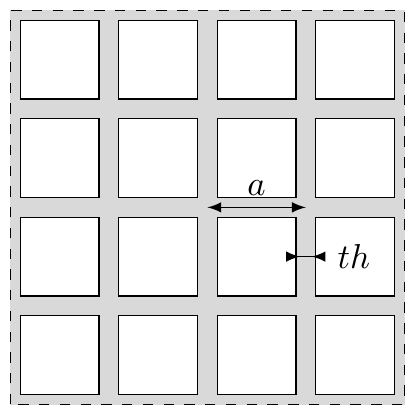}
\caption{$[\DD_4]$-invariant inner geometry}\label{fig:D4geom}
\end{figure}

This domain contains a material having an inner square architecture, as depicted on Figure \ref{fig:D4geom}. In terms of group language, the unit cell of this lattice is said to have a $[\DD_4]$ symmetry\footnote{$[\DD_{n}]$, refers to the dihedral group which is generated by a $n$-fold rotations and mirrors perpendicular to the rotation axis. $[\DD_{n}]$-invariant objects are achiral.}.
This architectured plate will be homogenized as a strain-gradient elastic continuum. The matrices associated to the homogenized constitutive law are extracted from \cite{ADR15}. For any orientation for which $\ii$ is collinear to a mirror lines of the architecture the matrix representations of the constitutive tensors have the following shape\footnote{Due to the $[\DD_{4}]$ symmetry there are two inequivalent orientation associated with this requirement. Consider a square, the first orientation corresponds to align $\ii$ with the line connecting the middle of opposite edges, while the second is associated with the line connecting opposite vertices. In these two orientations the shape  of the matrices will be identical, but not their values.}:
\begin{equation}
C_{\DD_{4}}=
\begin{pmatrix}
c_{11}&c_{12} &0    \\

      &c_{11} &0   \\
      
      &       &c_{33}    \\
\end{pmatrix}_\mathcal{B}\quad;\quad A_{\DD_{4}}=
\begin{pmatrix}
a_{11}&a_{12} &a_{13}     &0        &0    &0\\

      &a_{22} &a_{23}     &0  &0         &0\\
      
      &       &a_{33}    &0  &0   &0      \\
      
      &       &           &a_{11}   &a_{12}    &a_{13}\\
      
      &	 		  &           &         &a_{22}    &a_{23}\\
      
      &       &           &			    &          &a_{33}
\end{pmatrix}_\mathcal{B}\end{equation}
Since those matrices are symmetrical only half of each are defined. Solutions of strain-gradient elasticity for this anisotropic system have been studied in \cite{PAG16}. For a general displacement field:
\ben
\V{u}(\V{x})=u_1(x_{1},x_{2})\ii+u_2(x_{1},x_{2})\jj
\een
the PDE system associated to the bulk equation \eqref{eq:Bulk2}  is, in the static case,
\ben
\begin{cases}
c_{11}u_{1,11}+\tilde{c}_{12}u_{2,12}+\overline{c}_{33}u_{1,22}+f_1=a_{11}u_{1,1111}+\tilde{a}_{1}(u_{2,1112}+u_{2,1222})+\tilde{a}_{2}u_{1,1122}+\overline{a}_{33}u_{1,2222}\\
c_{11}u_{2,22}+\tilde{c}_{12}u_{1,12}+\overline{c}_{33}u_{2,11}+f_2=a_{11}u_{2,2222}+\tilde{a_{1}}(u_{1,1222}+u_{1,1112})+\tilde{a_{2}}u_{2,1122}+\overline{a}_{33}u_{2,1111}
\end{cases}
\een
where the following simplifications have been used
\begin{itemize}
\item $\overline{c}_{33}=\frac{1}{2}c_{33}$;
\item $\overline{a}_{33}=\frac{1}{2}a_{33}$;
\item $\tilde{c}_{12}=c_{12}+\frac{1}{2}c_{33}$;
\item  $\tilde{a}_{1}=a_{12}+\frac{\sqrt{2}}{2}(a_{13}+a_{23})+\frac{1}{2}a_{33}$;
\item  $\tilde{a}_{2}=a_{22}+\sqrt{2}(a_{13}+a_{23})+\frac{1}{2}a_{33}$.
\end{itemize}
Those equations have to be supplemented by  appropriate boundary conditions and depend on some specific combinations of 9 constitutive parameters $c_{11}$, $c_{12}$, $c_{33}$, $a_{11}$, $a_{13}$, $a_{23}$, $a_{22}$, $a_{12}$, $a_{33}$. 
In the following situation, which corresponds to the investigated one,  the PDE system is simplified:
\begin{enumerate}
\item the domain is finite along $\ii$ and infinite along $\jj$. As a consequence the domain is constituted of a unique row of square lattice and periodic boundary conditions are considered along edges $B$ and $D$ (c.f. Fig.\ref{bodyB});
\item Boundary conditions along edges $A$ and $C$ are independent of $x_{2}$.
\end{enumerate}
Under these hypotheses, the displacement field can be looked under the following form:
\ben
\V{u}(\V{x})=u_1(x_{1})\ii+u_2(x_{1})\jj
\een
and the PDE associate to bulk equilibrium reduce to:
\begin{equation}
\begin{cases}
c_{11}u_{1,11}-a_{11}u_{1,1111}+f_{1}=0\\
\overline{c}_{33}u_{2,11}-\overline{a}_{33}u_{2,1111}+f_2=0
\end{cases}
\label{PDEs1DSG}
\end{equation}
Supplemented by the boundary conditions:
\begin{equation}
\V{t}(0)=
\begin{pmatrix}
-\sigma_{11}(0)+\tau_{111,1}(0)=-c_{11}u_{1,1}(0)+a_{11}u_{1,111}(0)\\
-\sigma_{12}(0)+\tau_{121,1}(0)=-c_{33}u_{2,1}(0)+a_{33}u_{2,111}(0)\\
\end{pmatrix}
\quad ;\quad  
\V{R}(0)=
\begin{pmatrix}
-\tau_{111}(0)=-a_{11}u_{1,111}(0)\\
-\tau_{121}(0)=-a_{33}u_{2,111}(0)\\
\end{pmatrix},
\label{BC0}
\end{equation}
and
\begin{equation}
\V{t}(L)=
\begin{pmatrix}
\sigma_{11}(L)-\tau_{111,1}(L)=c_{11}u_{1,1}(L)-a_{11}u_{1,111}(L)\\
\sigma_{12}(L)-\tau_{121,1}(L)=c_{33}u_{2,1}(L)-a_{33}u_{2,111}(L)\\
\end{pmatrix}\quad ;\quad  
\V{R}(L)=
\begin{pmatrix}
\tau_{111}(L)=a_{11}u_{1,111}(L)\\
\tau_{121}(L)=a_{33}u_{2,111}(L)\\
\end{pmatrix}.
\label{BCL}
\end{equation}
The displacement field solution to the boundary value problem now  depends only on four material parameters: $c_{11}$, $\overline{c}_{33}$, $a_{11}$, $\overline{a}_{33}$. The aim of the static tests will be to identify those parameters. 

\subsection{Static identification}

To extract the strain-gradient elasticity parameters, 4 independent numerical experiments should be realized.
In Section \ref{Setexp_subsubsec} the general setting of that 4 experiments and in Section \ref{Numexp_subsubsec} their explicit definitions will be given.

\subsubsection{Settings of the experiments}
\label{Setexp_subsubsec}

\begin{figure}[H]
\centering
\includegraphics[scale=1]{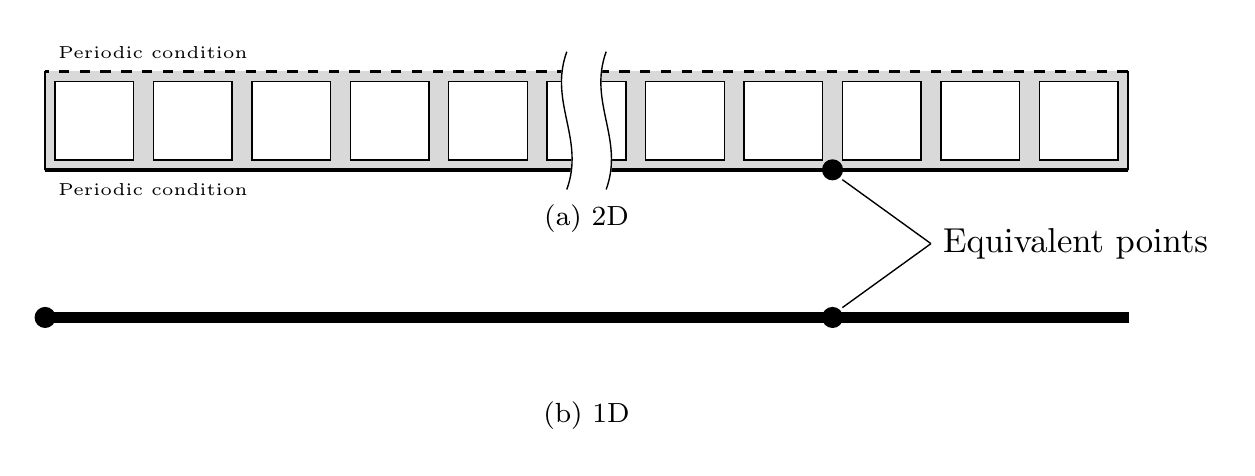}\\
\caption{Micro-macro identification.}\label{fig:micro-macro}
\end{figure}

The numerical experiments (Finite Element simulations) are conducted on a 2D architectured material with the assigned BCs indicated on Fig.\ref{fig:static}(a) and Fig.\ref{fig:static}(c), and whose properties are listed in Table.\ref{tab:param}).
Analytical solutions on a 1D SG material with assigned BCs, as indicated on Fig.\ref{fig:static}(b) and Fig.\ref{fig:static}(d), are determined. Next, the SG material parameters are estimated by fitting the numerical solution following the micro macro identification described in Fig. \ref{fig:micro-macro}.

Notations associated to the geometry description of the 1D strain-gradient elastic rod are represented in Fig. \ref{fig:static}b:
\begin{itemize}
\item $\Omega=]0;L[$;
\item $\partial\Omega=\{0\}\cup\{L\}$;
\item $\overline{\Omega}=\Omega\cup\partial\Omega$.
\end{itemize}

\begin{table}[H]
\centering
\begin{tabular}{ccccccccc}
\hline
 $a$   & $L$   & $\rho_m$  & $E_{m}$  & $\nu_{m}$  \\
  (cm) &  (cm) &(kg\;m$^{-3}$) &  (GPa) &  (-) \\
 \hline
1 &  100 &2000 & 200 & 0.3 \\
 \hline
\end{tabular}
\caption{Parameters used in the 2D numerical simulations. 
}\label{tab:param}
\end{table}
This identification procedure has been applied to different thicknesses $th$ (cf Fig.\ref{fig:D4geom}), and the results are summarized in Table.\ref{tab:coeffs}.

\subsubsection{Numerical experiments}
\label{Numexp_subsubsec}

The 4 experiments to be conducted are:
\begin{itemize}
\item 2 classical testings: Extension and Shear test;
\item 2 generalized testings: Hyper-Extension and Hyper-Shear test.
\end{itemize}

\begin{figure}[H]
\centering
\includegraphics[scale=1]{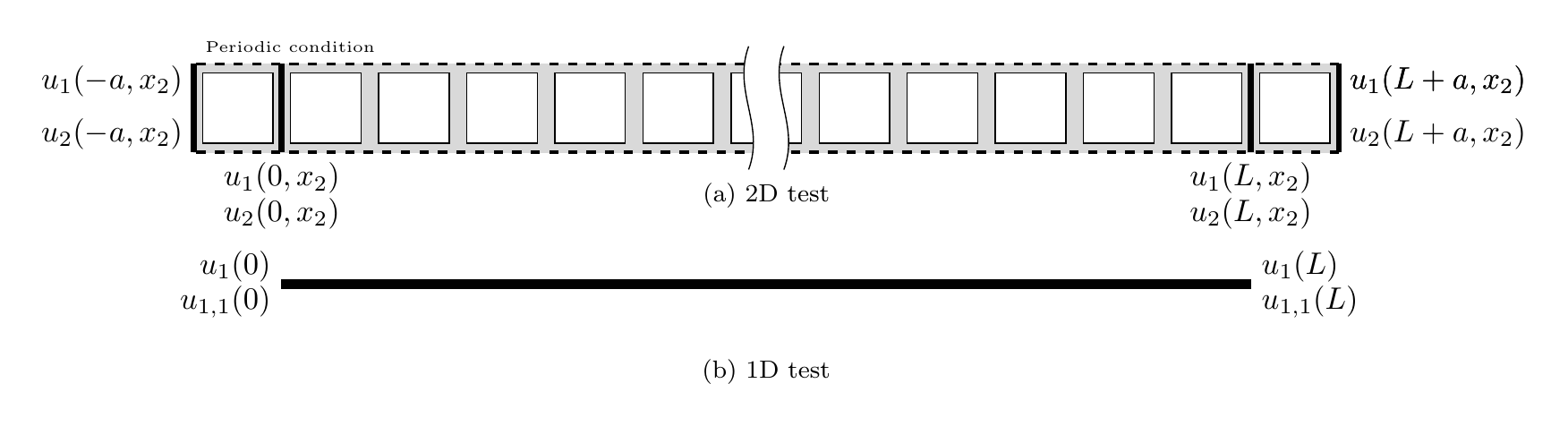}\\
\centering
\includegraphics[scale=1]{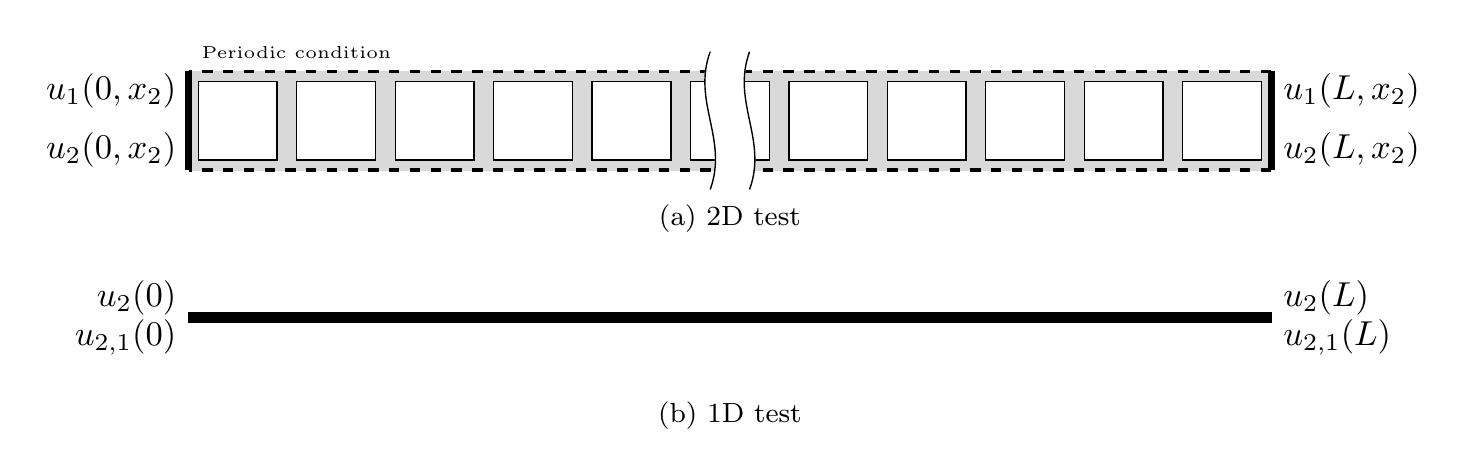}
\caption{Schematic representation of static (a) 2D and (b) 1D extension tests, (c) 2D and (d) 1D shear test. 
The 2D problems are numerically solved, while the 1D ones are analytically solved.}\label{fig:static}
\end{figure}

\noindent{\textbf{Extension test}}
The displacement field associated to the extension experiment (Fig.\ref{fig:static}b) is solution of the following ODE:
\begin{equation}\label{1Dx}
\begin{cases}
c_{11}u_{1,11}-a_{11}u_{1,1111}=0,\quad \forall x\in\Omega,\\
u_{1}(0)=0,\quad u_{1}(L)=\delta_{x},\\
R_{1}(0)=0,\quad R_{1}(L)=0
\end{cases}
\end{equation}
BCs correspond to prescribed horizontal displacement and free double force. 
By the constitutive law, the conditions of free double force, because of (\ref{BC0}) and (\ref{BCL}), on the boundaries are equivalent to: 
\ben
u_{1,11}(0)=0,\quad
u_{1,11}(L)=0
\een
The analytic solution to this boundary value problem (BVP) is:
\begin{equation} \label{solfiga}
u_{1}(x)=\dfrac{ \delta _x}{L}x.
\end{equation}
Thus, the traction $t_1$, at the right-hand side ($t_1(L)$) of the boundary $\partial\Omega$, because of (\ref{BC0}) and (\ref{BCL}), is
\begin{equation}
t_1(L)=c_{11}u'_{1}(L)=c_{11}\frac{\delta _x}{L}\Rightarrow c_{11}=t_1(L)\frac{L}{\delta _x}
\label{idexttestc11}
\end{equation}
The corresponding numerical test on the 2D structure, that is represented in Figure \ref{fig:static}a, is realized by imposing the following boundary conditions:
\begin{equation}
u_{1}(0,x_2)=0,\quad
u_{1}(L,x_2)=\delta_{x}
\label{BCext_micro}
\end{equation}
Partial differential equations are those for a standard Cauchy linearly elastic material with those material parameters exposed in Table \ref{tab:param}.
The reaction corresponding to the kinematic constraint (\ref{BCext_micro})$_2$ is identified with $t_1(L)$ and the identification of the material coefficient $c_{11}$ is done via eq. (\ref{idexttestc11}).
\noindent{\textbf{Hyper-Extension test}}
In this case the ordinary differential equation is the same as in the previous test, Eq. \eqref{1Dx}$_1$, but the boundary conditions are:
\begin{equation*}
u_{1}(0)=0,\quad
u_{1}(L)=\delta_{x},\quad
u_{1,1}(0)=0,\quad
u_{1,1}(L)=0,
\end{equation*}
that correspond to both displacement and gradient of displacement prescriptions.
Thus, the analytic solution, for the scheme represented in Fig.\ref{fig:static}b) is now more complicated:
\be
u_{1}(x)=\frac{\left(\text{Sinh} \left[\frac{L
   r_1}{2}\right]-\text{Sinh} \left[\frac{r_1 (L-2 x)}{2}\right]-xr_1 \text{Cosh} \left[\frac{L r_1}{2}\right]\right)\delta _x}{2 \text{Sinh} \left[\frac{L r_1}{2}\right]-L r_1 \text{Cosh} \left(\frac{L r_1}{2}\right)}\label{eq:an-H-Ext}
\ee
where the material parameter $r_1$ is defined as follows: $r_1=\sqrt{\frac{c11}{a11}}$.

The corresponding test on the 2D structure, Figure \ref{fig:static}a, is realized by imposing the following boundary conditions:
\begin{equation*}
u_{1}(-a,x_2)=0,\quad
u_{1}(0,x_2)=0,\quad
u_{1}(L,x_2)=\delta_{x},\quad
u_{1}(L+a,x_2)=\delta_{x},
\end{equation*}
where the normal gradient of the displacement has been prescribed by enlarging the domain in the $x_1$ direction to guarantee that at $x_1=0$ and at $x_1=L$ the normal gradient is zero as prescribed in the continuous model.
The value of the constitutive parameter $a_{11}$ is therefore determined by fitting the horizontal displacement field $u_1$ computed for the 2D model in the periodicity line with the analytic solution \eqref{eq:an-H-Ext}.\\

\noindent{\textbf{Shear test}}
The displacement field associated to the extension experiment (Fig.\ref{fig:static}b) is solution of:
\begin{equation}\label{1Dy}
\begin{cases}
\overline{c}_{33}u_{2,11}-\overline{a}_{33}u_{2,1111}=0\quad \forall x\in\Omega,\\
u_{2}(0)=0,\quad
u_{2}(L)=\delta_{x}\\
R_{2}(0)=0,\quad
R_{2}(L)=0
\end{cases}
\end{equation}
BCs correspond to those obtained with the prescription of the displacement and of free double force.

Using the constitutive law, because of (\ref{BC0}) and (\ref{BCL}), free double force on boundaries is equivalent to: 
\ben
u_{2,11}(0)=0,\quad
u_{2,11}(L)=0
\een
The analytic solution is therefore
\begin{equation} \label{solfigb}
u_{2}(x)=\dfrac{ \delta_y}{L}x.
\end{equation}
The corresponding test on the 2D structure, Figure \ref{fig:static}a, is realized by imposing the following boundary conditions:
\begin{equation*}
u_{2}(0,x_2)=0,\quad
u_{2}(L,x_2)=\delta_{x}.
\end{equation*}
As for the extension test, the constitutive coefficient (in this case $\overline{c}_{33}$) is computed using the traction $t_2$ at the right-hand side of the domain:
\ben
\overline{c}_{33}=t_2(L)\frac{L}{\delta_x}.
\een
\\
\noindent{\textbf{Hyper-Shear test}}
Even in this case the ordinary differential equation is the same as in the previous test, Eq. \eqref{1Dy}, and the boundary conditions are
\begin{equation*}
u_{2}(0)=0,\quad
u_{2}(L)=\delta_{x},\quad
u_{2,1}(0)=0,\quad
u_{2,1}(L)=0
\end{equation*}
that corresponds to the prescriptions of both displacement and gradient of displacement.
The solution is here given analytically,
\be
u_{2}(x)=\frac{\left(\text{Sinh} \left[\frac{L
   r_2}{2}\right]-\text{Sinh} \left[\frac{r_2 (L-2 x)}{2}\right]-xr_2\text{Cosh} \left[\frac{L r_2}{2}\right]\right)\delta_y}{2 \text{Sinh} \left[\frac{L r_2}{2}\right]-L r_2 \text{Cosh} \left(\frac{L r_2}{2}\right)}\label{eq:an-H-Sh}
\ee
with the material parameter $r_2=\sqrt{\overline{c}_{33}/\overline{a}_{33}}$.
The corresponding test on the 2D structure, Figure \ref{fig:static}c, is realized by imposing the following boundary conditions :
\begin{equation*}
u_{1}(0,x_2)=0,\quad
u_{2}(0,x_2)=0,\quad
u_{1}(L,x_2)=0,\quad
u_{2}(L,x_2)=\delta_{y}.
\end{equation*}
The value of the constitutive parameter $\overline{a}_{33}$ is determined by fitting the vertical displacement field $u_2$ computed for the 2D model in the periodicity line with the analytic solution \eqref{eq:an-H-Sh}.

\subsection{Dynamic identification}
The phase velocities computed from Equation \eqref{eq:EVP} are:
\ben
v_P(k,c_{11},\rho,a_{11},J_P)=\sqrt[]{\dfrac{c_{11}+a_{11}k^2}{\rho+J_P k^2}},\qquad
v_S(k,c_{33},\rho,a_{33},J_S)=\sqrt[]{\dfrac{c_{33}+a_{33}k^2}{\rho+J_S k^2}}
\een
These expressions are used in the minimization procedure (Equations \eqref{eq:opt}), considering the values of $\rho$, $c_{11}$, $c_{33}$, $a_{11}$, $a_{33}$ obtained from the static identification. The results of the procedure are resumed in Table \ref{tab:coeffs}. As can  be seen, effect of the micro inertia is more important for P-modes, while it is vanishing  for S-waves. Indeed, this effect can be related to the inertia of the vertical bars in  the microstructure.

\begin{figure}[H]%
\centering
\includegraphics[scale=1]{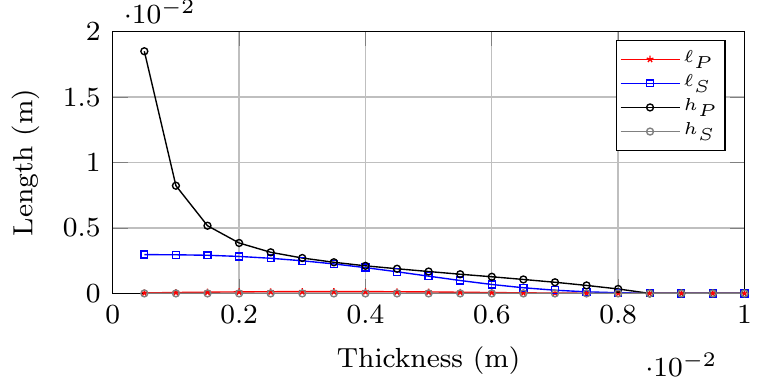}
\caption{Characteristic lengths for P- and S- waves}\label{fig:lengths}
\end{figure}

\begin{table}[H]
\centering
\begin{tabular}{c|ccccccccccc}
\hline
 $th$   & $\rho$ & $c_{11}$ & $\overline{c}_{33}$ & $a_{11}$  & $\overline{c}_{33}$  & $J_{P}$  & $J_{S}$  & $\ell_{P}$ & $\ell_{S}$  & $h_{P}$  & $h_{S}$ \\
  (mm) & (kg\;m$^-3$)& (MPa)& (MPa)& (Pa\;m$^2$) &  (Pa\;m$^2$) &  (kg\;m$^2$) & (kg\;m$^2$) &  (mm) &  (mm) &  (mm) &  (mm) \\
 \hline
 1. & 380. & 22.37 & 0.12 & 0.11 & 1.06 & 0.0257 & 0 & 0.07 & 2.95 & 20.16 & 0 \\
 2. & 720. & 45.79 & 1.09 & 0.62 & 8.71 & 0.0107 & 0 & 0.12 & 2.83 & 9.44 & 0 \\
 3. & 1020. & 70.67 & 4.1 & 1.36 & 25.53 & 0.0075 & 0 & 0.14 & 2.49 & 6.63 & 0 \\
 4. & 1280. & 97.6 & 10.63 & 1.85 & 41.3 & 0.0057 & 0 & 0.14 & 1.97 & 5.17 & 0 \\
 5. & 1500. & 127.54 & 21.74 & 1.57 & 37.8 & 0.0042 & 0 & 0.11 & 1.32 & 4.1 & 0 \\
 6. & 1680. & 161.46 & 36.72 & 0.7 & 17.16 & 0.0027 & 0 & 0.07 & 0.68 & 3.12 & 0 \\
 7. & 1820. & 198.3 & 52.55 & 0.14 & 3.18 & 0.0013 & 0 & 0.03 & 0.25 & 2.09 & 0 \\
 8. & 1920. & 233.52 & 65.75 & 0.01 & 0.17 & 0.0002 & 0 & 0.01 & 0.05 & 0.84 & 0 \\
 9. & 1980. & 259.61 & 74.13 & 0 & 0 & 0 & 0 & 0 & 0 & 0 & 0 \\
 10. & 2000. & 269.23 & 76.92 & 0 & 0 & 0 & 0 & 0 & 0 & 0 & 0 \\
 \hline
\end{tabular}
\caption{Table of the coefficients identified for different thicknesses. The size of the unit cell is $a=1$cm}\label{tab:coeffs}
\end{table}
The characteristic lengths of the strain-gradient model are defined as follows:
\ben
\ell_P=\sqrt{\dfrac{a_{11}}{c_{11}}},\quad
\ell_S=\sqrt{\dfrac{\overline{c}_{33}}{\overline{c}_{33}}},\quad
h_P=\sqrt{\dfrac{J_{P}}{\rho}},\quad
h_S=\sqrt{\dfrac{J_{S}}{\rho}}
\een
It is worth noting that these characteristic lengths depend on the geometry of the unit cell through the wall thickness.  This dependence is depicted on Figure \ref{fig:lengths}.

\begin{figure}[H]
\centering
\subfloat[Dispersion relations]{\includegraphics[width=0.4\linewidth]{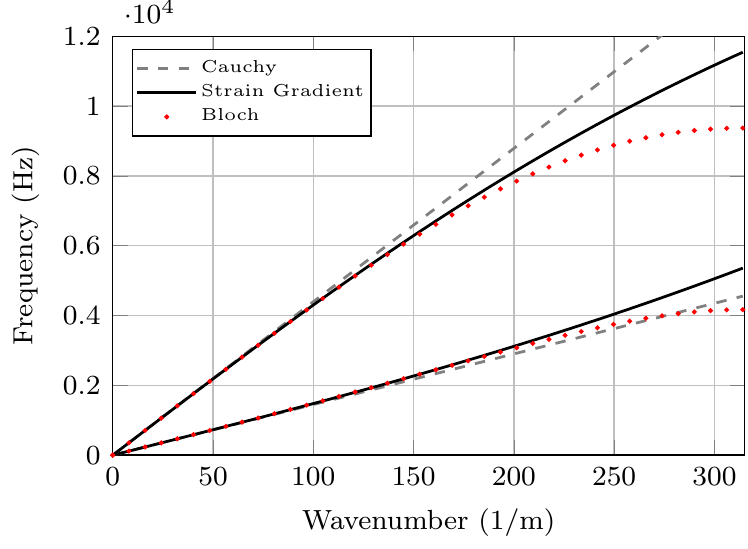}}\qquad
\subfloat[Phase velocity]{\includegraphics[width=0.4\linewidth]{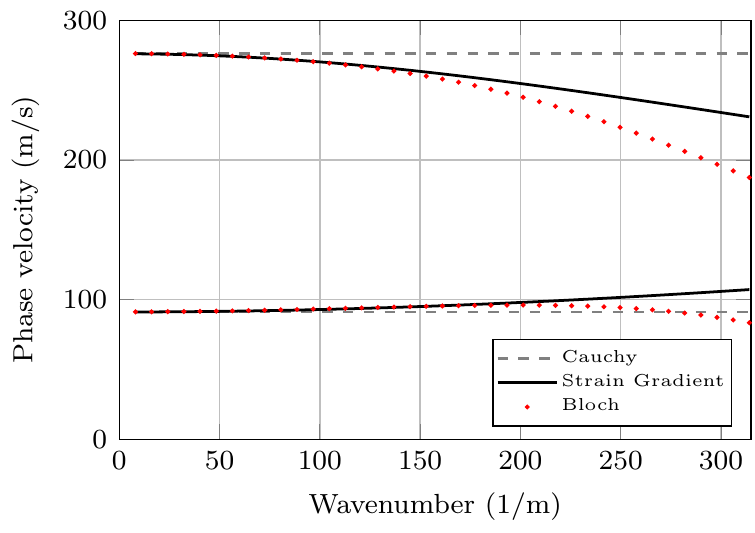}}\\
\subfloat[Group velocity]{\includegraphics[width=0.4\linewidth]{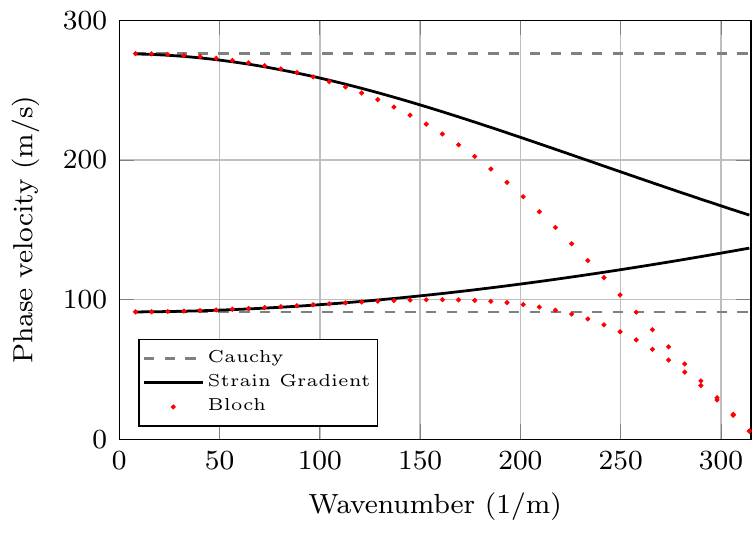}}
\caption{Dispersion curves, phase velocity and group velocity for the Strain gradient model (solid black lines), Cauchy model (dashed gray lines) compared with the results of FEM Bloch computation (black points) in the case of $a=1$cm and $th=4$mm.}\label{fig:disp}
\end{figure}

In Figure \ref{fig:disp} we can observe the result of the identification procedure by plotting the superposition of the dispersion curves, phase velocity and group velocity  for the strain-gradient model (solid lines) and the FEM model (points). The result of the identification procedure for unit cell having wall thickness of for $th=4$mm is plotted on Figure \ref{fig:disp}. On associated subfigures the dispersion curves, phase velocity and group velocity for the strain-gradient model (solid lines), Cauchy model (dashed gray lined) and the FEM model (black points) have been drawn. As can be observed, a good fit is achieved in the first third of the Brillouin zone, while the model clearly loose accuracy for high values of $k$. As previously discussed, it can be observed that group velocity is diverging faster than the other quantities. This illustrate the fact that the quality of the approximation should not only be assessed on the accurate description of the dispersion curves.

\subsection{Estimation of the error}
\begin{figure}[H]
\centering
\includegraphics[width=0.8\linewidth]{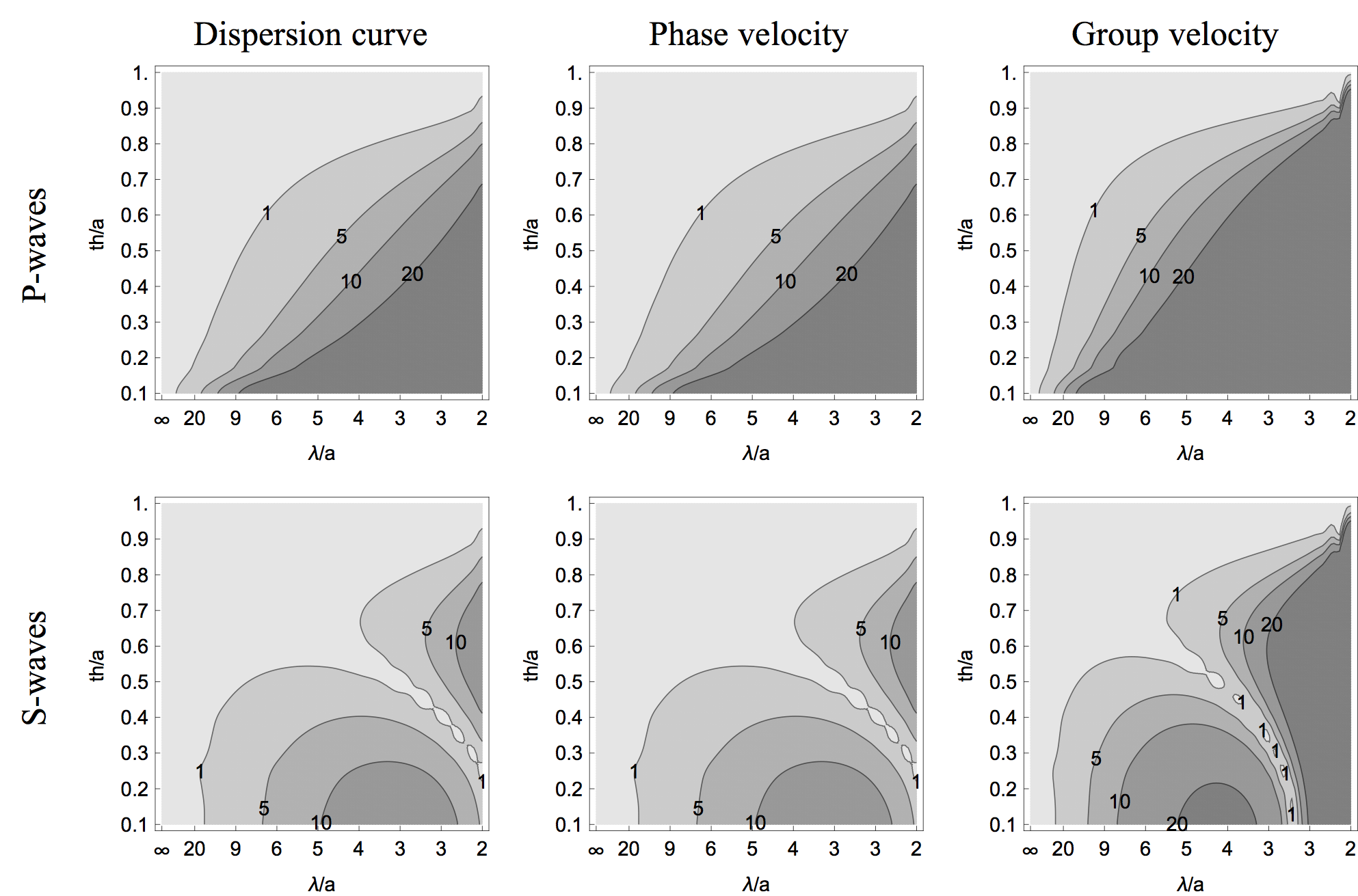}
\caption{Contour plots of the error  (in \%) between the Cauchy overall solution and the Bloch analysis for the dispersion curves, phase velocity and group velocity.}\label{fig:err1G}
\end{figure}

\begin{figure}[H]
\centering
\includegraphics[width=0.8\linewidth]{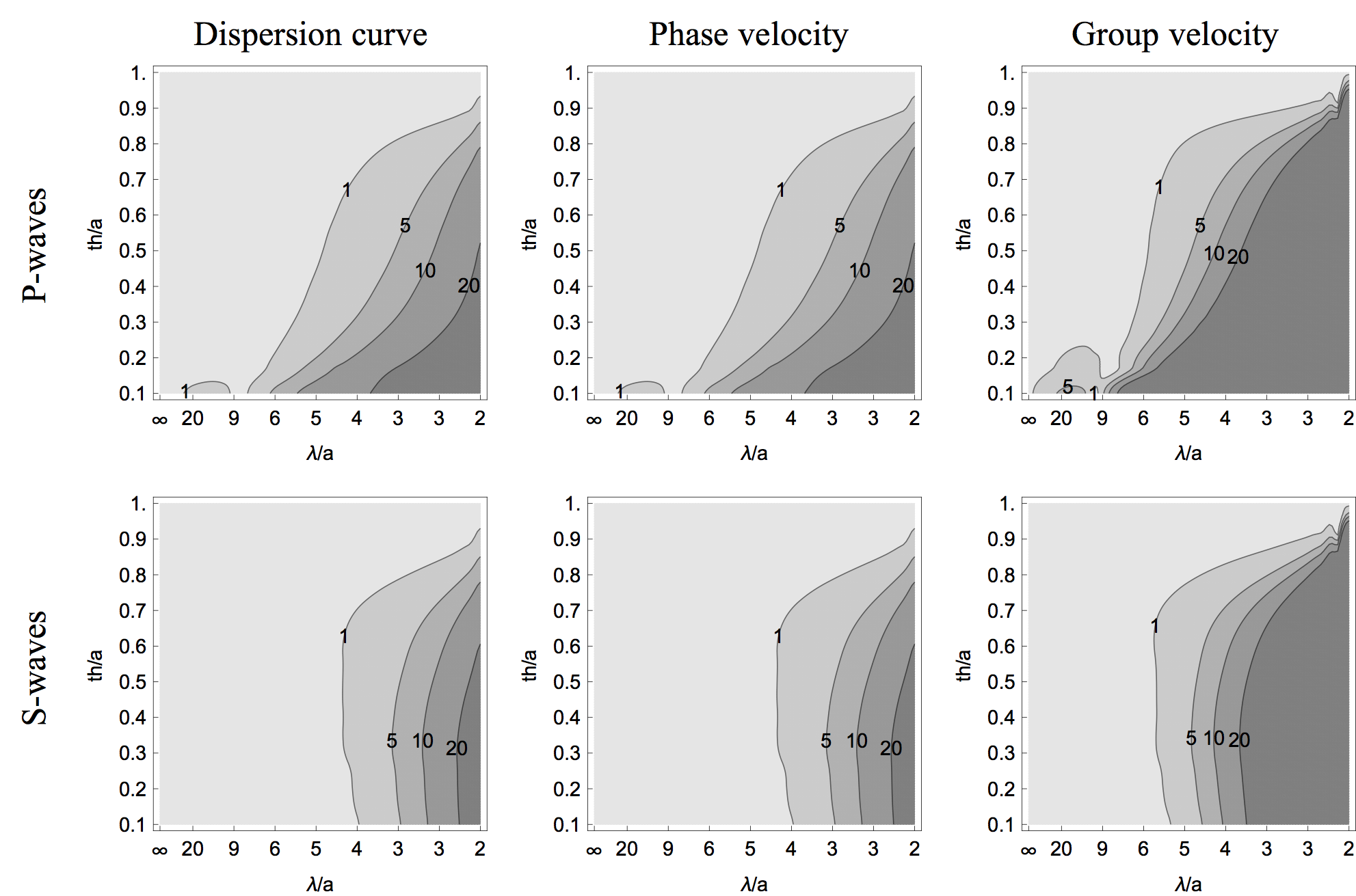}
\caption{Error  (in \%) between the SGE overall solution and the Bloch analysis for the dispersion curves, phase velocity and group velocity.}\label{fig:err2G}
\end{figure}

In this section we estimate the error between the FEM computation (Bloch analysis) and the overall strain-gradient model. Results are then compared  with a classical Cauchy overall continuum. 
This error computation will be performed for different thicknesses $th$ of the  microstructure walls.  The abscissa of each subplot represent the wavelength $\lambda$ over the size of the unit cell $a$, while the ordinates represent the thickness of the walls $th$ with respect to $a$.  This means that in the upper part of each subplot, where $th/a=1$, the microstructure is completely homogeneous, while in the lower part the walls are very thin. In the case of  $th/a=1$, as the medium is not dispersive we do expect that both model perform correctly.
In figures \ref{fig:err1G} and \ref{fig:err2G} the first row represent the P-waves, the second raw the S-waves. The left, the central and the right columns, represent the errors in terms of dispersion, phase velocity and group velocity, respectively. By comparing the errors shown in Figure \ref{fig:err1G} and \ref{fig:err2G} for classical and strain-gradient approximation, respectively, we can see that the strain-gradient model behaves  better in the case of thin walls, down to  values of wavelength around six times the size of the unit cell. In this zone, the SG model has less than 1\% error for almost all configurations, while the  first gradient model has, in the same zones up to 10-20\% errors. The first gradient plots for S-waves have a slightly better behavior in a narrow zone for lower values of $\lambda/a$. But this is only due to the fact that the dispersion curves, as well as the phase and group velocity, change concavity and cross the first gradient curves, as it can be observed in Figure \ref{fig:disp}.

Another remark can be made about the decrease of performances of the strain-gradient model when the ratio $th/a$ becomes very small. 

\subsection{Time domain validation}

In this last subsection we compare the transient time domain response (\textit{i.e.} we solve the boundary value problem) of a 2D FEM  simulations with the equivalent homogeneous 1D strain-gradient model (add figure) whose equations of motion in the $[\DD_4]$ case are the following:
\begin{figure}[H]
\centering
\includegraphics[width=0.8\linewidth]{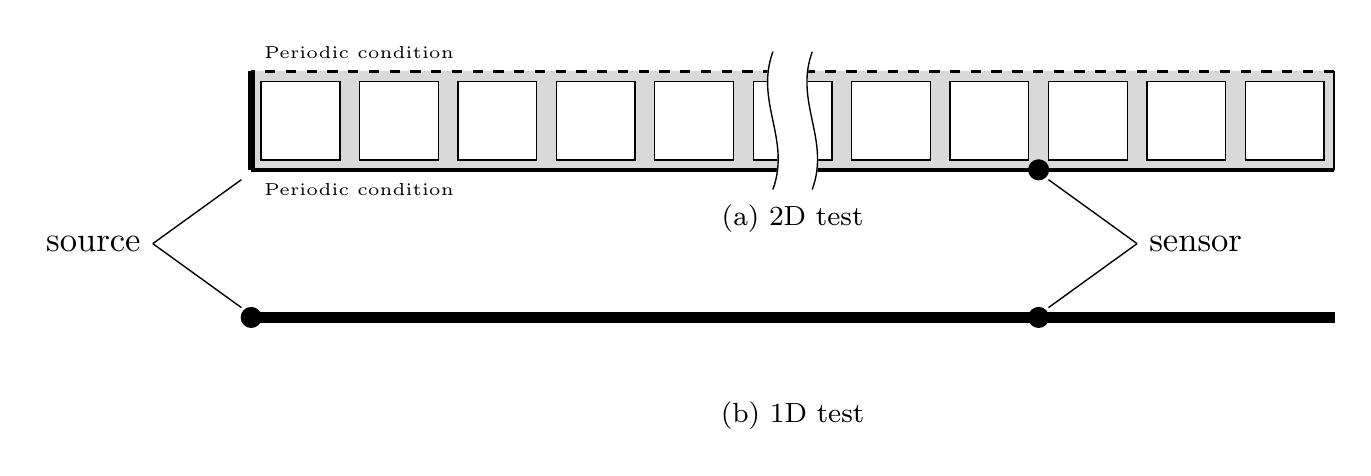}
\caption{Schematic description of the time domain test.}\label{fig:TD-config}
\end{figure}
\ben
\begin{cases}
c_{11}u_{1,11}-a_{11}u_{1,1111}=\rho \ddot{u}_1-J_P \ddot{u}_{1,11}\\
\overline{c}_{33}u_{2,11}-\overline{a}_{33}u_{2,1111}=\rho \ddot{u}_2-J_S \ddot{u}_{1,22}
\end{cases}
\een


The numerical configuration is the following: a displacement source placed at $x=0$ is exciting a transient wave with a given central frequency $f_c$. The  central frequency for each test, as well as the corresponding unit cells per wavelength ratio, are resumed in Table \ref{tab:freqs}.
\begin{table}[H]
\centering
\begin{tabular}{cc|cc}
\hline
\multicolumn{2}{c|}{P-waves} & \multicolumn{2}{|c}{S-waves}\\
\hline
$f_c$ (kHz) & $\lambda/a $(-) & $f_c$  (kHz) & $\lambda/a$ (-)\\
\hline
1  & 27.5 & 0.2  & 45.6\\
4 & 6.8  & 1.5 & 6.2\\
\hline
\end{tabular}
\caption{Central frequencies and corresponding unit cells per wavelength ratio.}\label{tab:freqs}
\end{table}

\begin{figure}%
\centering
\subfloat[ 1 kHz ]{\includegraphics[width=0.4\linewidth]{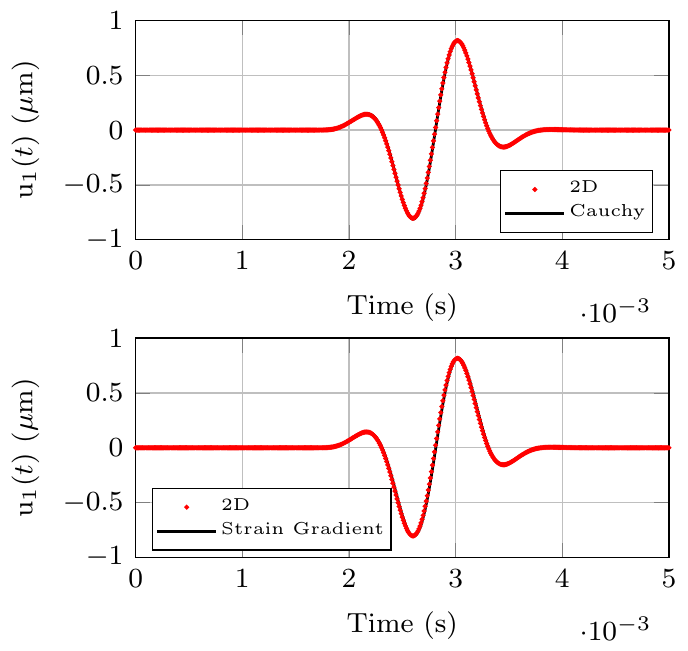}}
\subfloat[ 4 kHz ]{\includegraphics[width=0.4\linewidth]{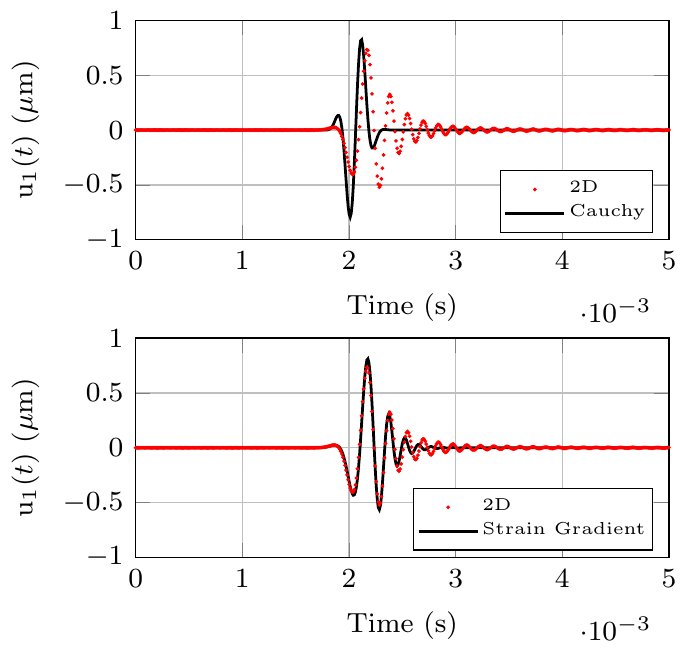}}\\
\caption{Time domain response P-waves evaluated at $x=50$ cm.}\label{fig:P-t}
\end{figure}

\begin{figure}%
\centering
\subfloat[ 0.2 kHz ]{\includegraphics[width=0.4\linewidth]{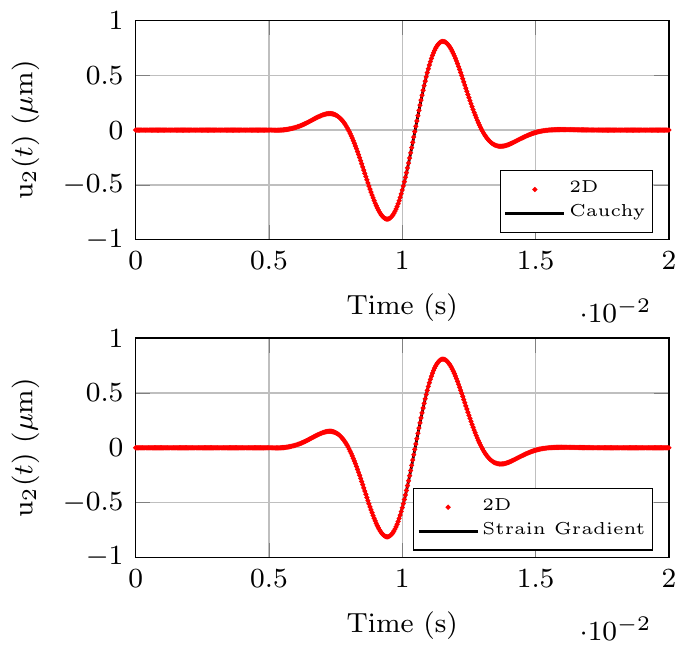}}
\subfloat[ 1.5 kHz ]{\includegraphics[width=0.4\linewidth]{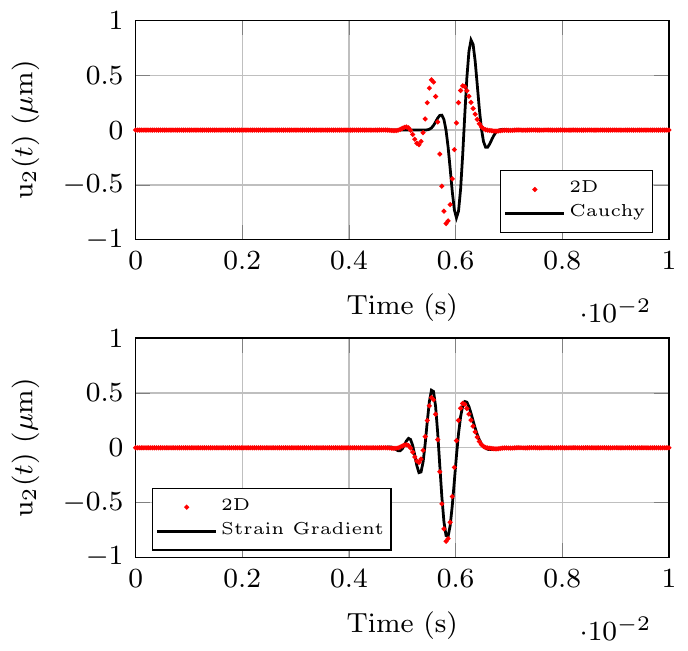}}\\
\caption{Time domain response S-waves evaluated at $x=50$ cm.}\label{fig:S-t}
\end{figure}

\begin{figure}%
\centering
\subfloat[  ]{\includegraphics[width=0.4\linewidth]{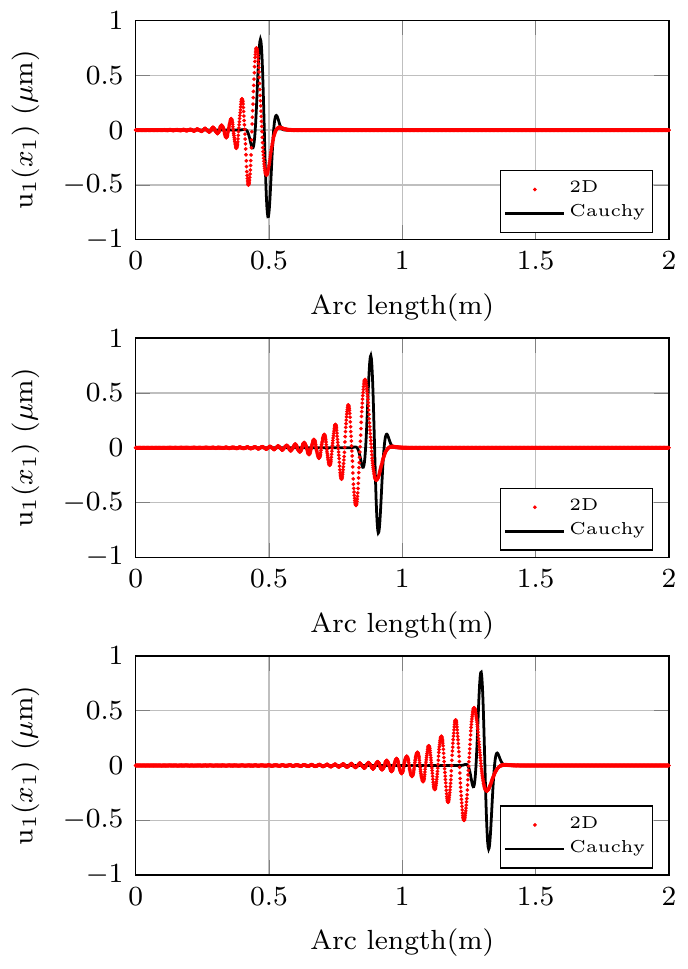}}
\subfloat[ ]{\includegraphics[width=0.4\linewidth]{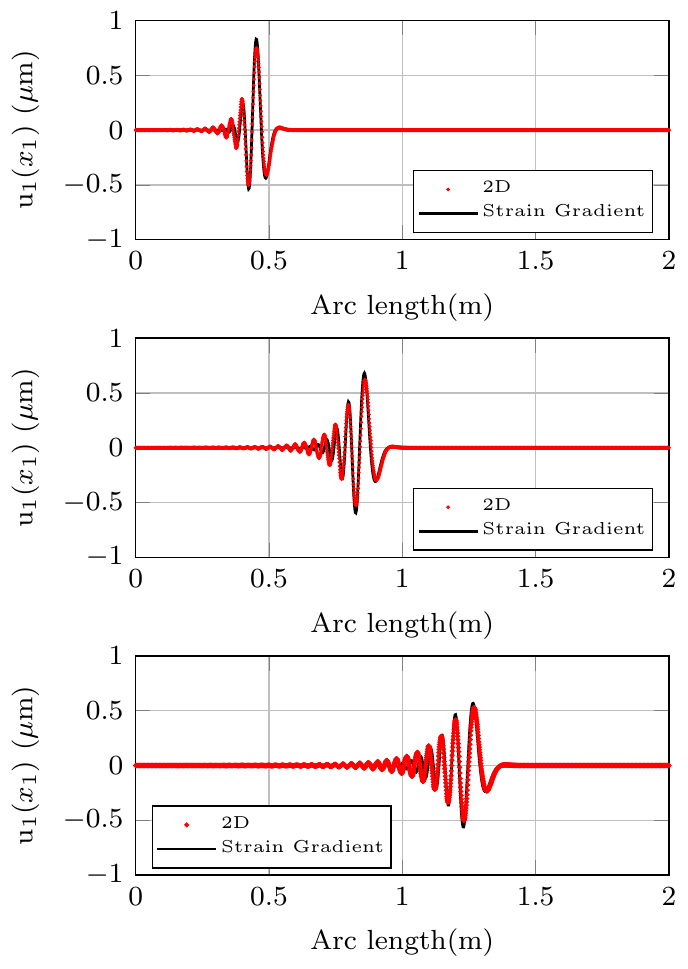}}
\caption{Longitudinal cut at different time instants for P-waves or $f_c=4$ kHz and for $th=4$mm.}\label{fig:P-x}
\end{figure}

\begin{figure}%
\centering
\subfloat[  ]{\includegraphics[width=0.4\linewidth]{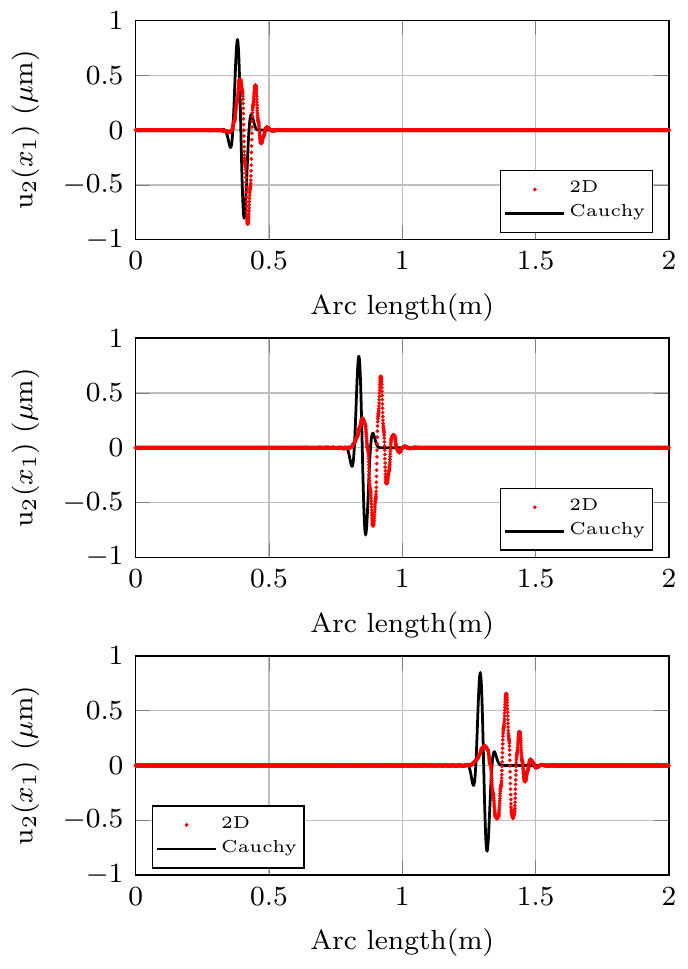}}
\subfloat[ ]{\includegraphics[width=0.4\linewidth]{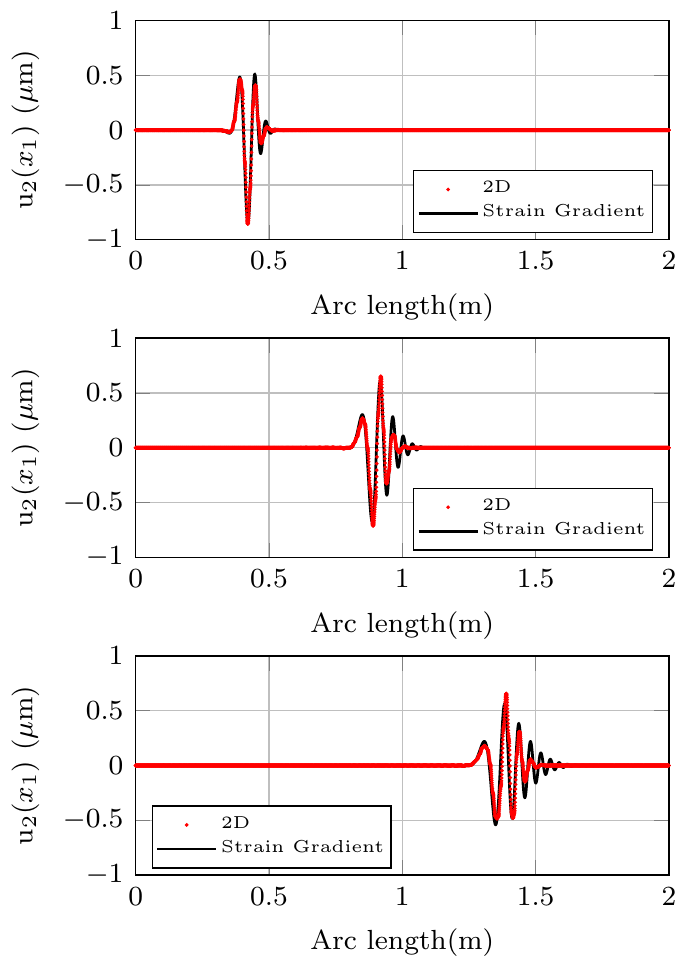}}
\caption{Longitudinal cut at different time instants for S-waves for $f_c=1.2$ kHz and for $th=4$mm.}\label{fig:S-x}
\end{figure}

The results are presented in two forms:
\begin{itemize}
\item displacement field measured at $x=50$ cm (Figures \ref{fig:P-t} and \ref{fig:S-t});
\item displacement field on a longitudinal cut line (symmetry line) of the 2D model compared to the result for the 1D model (Figures \ref{fig:P-x} and \ref{fig:S-x}.).
\end{itemize}
A schematic description of the tests is presented in Figure \label{fig:TD-config}.
The results show that a quantitative agreement is observed with the strain-gradient model and on the contrary a Cauchy model looses accuracy whenever the frequency is raised.
In particular, Figure \ref{fig:P-t} shows that the strain-gradient model is able to account for back scattering effects. This effect is due to the fact that, as it can be seen in Figure \ref{fig:disp} and \ref{fig:P-x}, small wavelengths travel slower in the case of P-waves. The opposite effect can be remarked for S-waves, where the so called anomalous dispersion (\textit{i.e.} group velocity higher than phase velocity) can be observed. In the case of normal dispersion (Figure \ref{fig:P-x}) the Cauchy solution is aligned with the head of the traveling signal, while for anomalous dispersion it is aligned with the tail (Figure \ref{fig:S-x}). This latter remark can be important when considering the velocity of the first arriving signal, used in quantitative ultrasonic characterization, as in \cite{Rosi:2015ci}. 

\section{Conclusions}\label{s:Con}
In this paper we presented a quantitative estimation of the validity domain of the overall description of wave propagation by the strain gradient elasticity model.
The results are obtained for a unidirectional wave propagation, but the extension to multidirectional situations is possible. The challenge for such an extension, is based on the estimation of the overall coefficients necessary to set up the model.

The main innovative results of the paper are the following:
\begin{itemize}
\item A novel mixed static-dynamic identification procedure for the identification of the coefficients is introduced. The procedure involves a static analysis for the identification of the coefficients of the first and second order elastic tensors (i.e $\qT{C}$ and $\sT{A}$) based on structural type tests ans a dynamic analysis based on Bloch theorem for the identification of the microinertia tensor $\qT{J}$.

\item the identification procedure is applied to a pseudo-1D case involving a microstructure with $[\DD_4]$-invariant inner geometry, with the wave-vector parallel to a main direction of symmetry.

\item the domain of validity is evaluated in term of wavelength. For the present case, and considering a maximum error of 1\%, the limit of the SGE model is evaluated at a ratio wavelength over size of the microstructure equal to six. It should be note that in the same situation the limit of the Cauchy elasticity is evaluated at a ratio wavelength over size of the microstructure equal to 20.

\item the time domain transient response of the strain gradient elasticity model is compared with the response of the 2D plane strain finite element computation of the actual structure. Within the validity range of the model, the solution of strain gradient elasticity fits very well the 2D solution while a standard Cauchy description loses accuracy as the wavelength decreases.
\end{itemize}

\bibliographystyle{apalike}
\bibliography{Bib}

\begin{thebibliography}{}

\bibitem[Askes and Aifantis, 2006]{AA06}
Askes, H. and Aifantis, E. (2006).
\newblock Gradient elasticity theories in statics and dynamics-a unification of
  approaches.
\newblock {\em International Journal of Fracture}, 139(2):297--304.

\bibitem[Auffray et~al., 2015]{ADR15}
Auffray, N., Dirrenberger, J., and Rosi, G. (2015).
\newblock A complete description of bi-dimensional anisotropic strain-gradient
  elasticity.
\newblock {\em International Journal of Solids and Structures}, 69:195--206.

\bibitem[Auffray et~al., 2016]{AKO16}
Auffray, N., Kolev, B., and Olive, M. (2016).
\newblock Handbook of bi-dimensional tensors: Part i: Harmonic decomposition
  and symmetry classes.
\newblock {\em Mathematics and Mechanics of Solids}, page 1081286516649017.

\bibitem[Bacigalupo and Gambarotta, 2014]{BG14b}
Bacigalupo, A. and Gambarotta, L. (2014).
\newblock {Second-gradient homogenized model for wave propagation in
  heterogeneous periodic media}.
\newblock {\em International Journal of Solids and Structures},
  51(5):1052--1065.

\bibitem[Berezovski et~al., 2011]{BEB11}
Berezovski, A., Engelbrecht, J., and Berezovski, M. (2011).
\newblock Waves in microstructured solids: a unified viewpoint of modeling.
\newblock {\em Acta mechanica}, 220(1-4):349--363.

\bibitem[Brillouin, 2003]{Bri03}
Brillouin, L. (2003).
\newblock {\em Wave propagation in periodic structures: electric filters and
  crystal lattices}.
\newblock Courier Corporation.

\bibitem[Chen et~al., 2014]{CLH14}
Chen, Y., Liu, X., and Hu, G. (2014).
\newblock Micropolar modeling of planar orthotropic rectangular chiral
  lattices.
\newblock {\em Comptes Rendus M{\'e}canique}, 342(5):273--283.

\bibitem[Cosserat and Cosserat, 1909]{Cos09}
Cosserat, E. and Cosserat, F. (1909).
\newblock {\em Théorie des corps déformables}.
\newblock Paris.

\bibitem[dell'Isola et~al., 2014]{dell2014complete}
dell'Isola, F., Maier, G., Perego, U., Andreaus, U., Esposito, R., and Forest,
  S. (2014).
\newblock The complete works of gabrio piola: volume i.
\newblock {\em Cham, Switzerland: Springer}.

\bibitem[dell’Isola et~al., 2015]{dell2015origins}
dell’Isola, F., Andreaus, U., and Placidi, L. (2015).
\newblock At the origins and in the vanguard of peridynamics, non-local and
  higher-gradient continuum mechanics: An underestimated and still topical
  contribution of {G}abrio {P}iola.
\newblock {\em Mathematics and Mechanics of Solids}, 20(8):887--928.

\bibitem[Dresselhaus et~al., 2007]{DDJ08}
Dresselhaus, M.-S., Dresselhaus, G., and Jorio, A. (2007).
\newblock {\em Group theory: application to the physics of condensed matter}.
\newblock Springer Science \& Business Media.

\bibitem[Erigen, 1967]{Eri68}
Erigen, A. (1967).
\newblock Theory of micropolar elasticity.
\newblock In Leibowitz, H., editor, {\em Fracture, vol. 2.}, pages 621--629.
  Academic Press, New York.

\bibitem[Farzbod and Leamy, 2011]{FL11}
Farzbod, F. and Leamy, M.~J. (2011).
\newblock Analysis of bloch's method and the propagation technique in periodic
  structures.
\newblock {\em Journal of vibration and acoustics}, 133(3):031010.

\bibitem[Gazalet et~al., 2013]{GDK+13}
Gazalet, J., Dupont, S., Kastelik, J.-C., Rolland, Q., and Djafari-Rouhani, B.
  (2013).
\newblock A tutorial survey on waves propagating in periodic media: Electronic,
  photonic and phononic crystals. perception of the bloch theorem in both real
  and fourier domains.
\newblock {\em Wave Motion}, 50(3):619--654.

\bibitem[Germain, 1973]{Ger73}
Germain, P. (1973).
\newblock The method of virtual power in continuum mechanics. part 2:
  Microstructure.
\newblock {\em SIAM Journal on Applied Mathematics}, 25(3):556--575.

\bibitem[Green and Rivlin, 1964]{GR64}
Green, A. and Rivlin, R. (1964).
\newblock Multipolar continuum mechanics.
\newblock {\em Archive for Rational Mechanics and Analysis}, 17(2):113--147.

\bibitem[Liu et~al., 2012]{LHH12}
Liu, X., Huang, G., and Hu, G. (2012).
\newblock Chiral effect in plane isotropic micropolar elasticity and its
  application to chiral lattices.
\newblock {\em Journal of the Mechanics and Physics of Solids},
  60(11):1907--1921.

\bibitem[Metrikine, 2006]{Met06}
Metrikine, A.~V. (2006).
\newblock On causality of the gradient elasticity models.
\newblock {\em Journal of Sound and Vibration}, 297(3):727--742.

\bibitem[Mindlin, 1964]{Min64}
Mindlin, R. (1964).
\newblock Micro-structure in linear elasticity.
\newblock {\em Archive for Rational Mechanics and Analysis}, 16(1):51--78.

\bibitem[Mindlin, 1965]{Min65}
Mindlin, R. (1965).
\newblock Second gradient of strain and surface-tension in linear elasticity.
\newblock {\em International Journal of Solids and Structures}, 1(4):417--438.

\bibitem[Mindlin and Eshel, 1968]{ME68}
Mindlin, R. and Eshel, N. (1968).
\newblock On first strain-gradient theories in linear elasticity.
\newblock {\em International Journal of Solids and Structures}, 4(1):109--124.

\bibitem[Nassar et~al., 2015a]{NHA15c}
Nassar, H., He, Q.-C., and Auffray, N. (2015a).
\newblock {A generalized theory of elastodynamic homogenization for periodic
  media}.
\newblock {\em Proceedings of the Royal Society A}, page Submitted.

\bibitem[Nassar et~al., 2015b]{NHA15b}
Nassar, H., He, Q.-C., and Auffray, N. (2015b).
\newblock {On asymptotic elastodynamic homogenization approaches for periodic
  media}.
\newblock {\em Journal of Mechanics and Physics of Solids}, 88.

\bibitem[Nassar et~al., 2015c]{NHA15a}
Nassar, H., He, Q.-C., and Auffray, N. (2015c).
\newblock {Willis elastodynamic homogenization theory revisited for periodic
  media}.
\newblock {\em Journal of the Mechanics and Physics of Solids}, 77:158--178.

\bibitem[Neff et~al., 2014]{NGM14}
Neff, P., Ghiba, I.-D., Madeo, A., Placidi, L., and Rosi, G. (2014).
\newblock A unifying perspective: the relaxed linear micromorphic continuum.
\newblock {\em Continuum Mechanics and Thermodynamics}, 26(5):639--681.

\bibitem[Norris and Shuvalov, 2011]{NS11}
Norris, A. and Shuvalov, A. (2011).
\newblock {Elastic cloaking theory}.
\newblock {\em Wave Motion}, 48(6):525--538.

\bibitem[Olive and Auffray, 2014]{OA14}
Olive, M. and Auffray, N. (2014).
\newblock Symmetry classes for odd-order tensors.
\newblock {\em ZAMM-Journal of Applied Mathematics and Mechanics},
  94(5):421--447.

\bibitem[Placidi et~al., 2015]{PAD+15}
Placidi, L., Andreaus, U., Della~Corte, A., and Lekszycki, T. (2015).
\newblock Gedanken experiments for the determination of two-dimensional linear
  second gradient elasticity coefficients.
\newblock {\em Zeitschrift f{\"u}r angewandte Mathematik und Physik},
  66(6):3699--3725.

\bibitem[Placidi et~al., 2016]{PAG16}
Placidi, L., Andreaus, U., and Giorgio, I. (2016).
\newblock Identification of two-dimensional pantographic structure via a linear
  d4 orthotropic second gradient elastic model.
\newblock {\em Journal of Engineering Mathematics}, 103:1--21.

\bibitem[Rosi and Auffray, 2016]{RA16}
Rosi, G. and Auffray, N. (2016).
\newblock Anisotropic and dispersive wave propagation within strain-gradient
  framework.
\newblock {\em Wave Motion}, 63:120--134.

\bibitem[Rosi et~al., 2016]{Rosi:2015ci}
Rosi, G., Nguyen, V.-H., and Naili, S. (2016).
\newblock {Numerical investigations of ultrasound wave propagating in long
  bones using a poroelastic model}.
\newblock {\em Mathematics and Mechanics of Solids}, 21(1):119--133.

\bibitem[Toupin, 1962]{Tou62}
Toupin, R. (1962).
\newblock Elastic materials with couple-stresses.
\newblock {\em Archive for Rational Mechanics and Analysis}, 11(1):385--414.

\bibitem[Willis, 1985]{Wil85}
Willis, J. (1985).
\newblock {The nonlocal influence of density variations in a composite}.
\newblock {\em International Journal of Solids and Structures}, 21(7):805--817.

\bibitem[Willis, 1997]{Wil97}
Willis, J. (1997).
\newblock {Dynamics of composites}.
\newblock In Suquet, P., editor, {\em Continuum Micromechanics}, chapter
  Dynamics o, pages 265--290. Springer-Verlag, New York.

\end{thebibliography}

\end{document}